\documentclass{aa}
\usepackage{epsfig}
\begin{document}

\title{Cool carbon stars in the halo II: a study of 25 new objects
\thanks{Based on observations made at the European Southern Observatory, Chile
 (program  73.A-0030), and on data from the
2MASS project (University of Massachusetts and IPAC/Caltech, USA).}}

\author{N. Mauron\inst{1} \and  
T.R. Kendall\inst{2}\thanks{Present address: Centre for Astrophysics 
Research, Science \& Technology 
Research Institute, University of Hertforshire, College Lane, 
Hatfield AL10 9AB, United Kingdom}
 \and K. Gigoyan\inst{3}}
\offprints{N.Mauron}
\institute{ Groupe d'Astrophysique, UMR 5024 CNRS, Case CC72,
 Place Bataillon, F-34095 Montpellier Cedex 5, France\\
 \email{mauron@graal.univ-montp2.fr}
  \and
 Laboratoire d'Astrophysique, Observatoire de Grenoble, Universit\'e
Joseph Fourier, BP 53, 38041 Grenoble Cedex 9, France
 \and
378433 Byurakan Astrophysical Observatory \& Isaac
 Newton Institute of Chile, Armenian Branch, Ashtarak d-ct, Armenia}

\date{Received xxx/  Accepted xxx}

\abstract{We present new results from an ongoing survey of carbon-rich
asymptotic giant (AGB) stars  in the halo of our Galaxy. After selecting candidates
primarily through their 2MASS colours, slit spectroscopy was achieved  at
the ESO NTT telescope. Twenty-one new AGB carbon stars
were discovered, increasing the total of presently known similar 
AGB C stars to $\sim$ 120. 
A further four were observed again in order to confirm their carbon-rich nature
and measure radial velocities. 
Two main findings  emerge from this work. First, we found 
a C star located at $\approx$\,130\,kpc
from the Sun  and at $b = -62^{\circ}$. This
distant star is remarkably close ($5$\,kpc) to the principal plane of the Stream of the
Sagittarius dwarf galaxy, and is likely to be a tracer of a distant poorly populated 
southern warp of the Stream. Such a warp is predicted by model simulations,
but it passes $\sim$\,45\,kpc from that star.
The second result is  that, mainly in the North,
several already known or newly discovered AGB
carbon stars lie far, up to 60\,kpc, from the mean plane of  the Sagittarius Stream.
  }

\titlerunning{Halo carbon stars}
\authorrunning{ N.Mauron, T.R. Kendall, K. Gigoyan}
\maketitle
% ------------------------------------------------------------------
\section{Introduction}

This paper is a continuation of our previous paper devoted to cool, asymptotic
giant branch (AGB), carbon (C) stars that are found at high galactic latitude
or, alternatively, far from the galactic plane  (Mauron et al. \cite{mauron04}, hereafter
Paper~I).  Historically, these stars have attracted the attention of observers
for several reasons: their characteristic observability due to the 
CN and C$_{\tiny2}$ bands in their spectra, their rarity ($\sim$ 100 are presently
 known), and their high luminosity as AGB stars, which make them  useful
for studying the  stellar content of the galactic halo at large distances
(see for example Wallerstein \& Knapp \cite{wallerstein98};  Totten \& Irwin \cite{totten98}, 
hereafter TI98).
For illustration, in the $I$-band, Demers et al. (\cite{demers03})  found from the study of
several Local Group galaxies that  the average $I$-band absolute
magnitude  of (not too dusty) AGB C stars 
is $M_{\tiny I}$\,=\,$-4.6$. This average luminosity leads one to expect
an apparent magnitude as bright as $I$\,=\,15.4  for a distance of 100\,kpc.

In the halo of the Galaxy, the  majority, but not all, of the cool AGB C stars
originate in the debris of the Sagittarius (Sgr) dwarf galaxy that orbits
the Milky Way (Ibata et al. \cite{ibata01}; see also Paper~I). 
One could also think that  the Magellanic Stream might contain
carbon stars, but no stellar population has been discovered in this stream 
despite many investigations, such as deep imagery
with the Keck telescope (Guhathakurta \& Reitzel \cite{guhathakurta98},
 and references therein).

 The cool  stars that we study here are
different from, and rarer than, other types of C rich stars
found with deep observations at high galactic latitude,
e.g. the high velocity CH stars and other types of warm C rich metal poor
giants, and the population of C rich dwarfs
(for more details see e.g., Wallerstein \& Knapp \cite{wallerstein98}, Margon et al.
\cite{margon02}, Downes et al. \cite{downes04}). 
In this paper,  we focus on the AGB-type stars,
which are usually understood  as originating in intermediate age populations. 
The important
point that drives our work is that the census of
these cool and luminous AGB  C stars located out of the galactic plane
is far from complete. Our goal is therefore to improve
our knowledge of this population of halo C stars
and to increase the number of these objects.

 A first step  towards achieving this task is to obtain a selection  of candidates
located at latitudes $|b| > 30^\circ$ and faint enough to avoid members
of the galactic disk. Subsequently, one has to provide  confirmation  with follow-up
spectroscopy. This experiment related in Paper~I  showed that about one third of
the candidates  selected with 2MASS photometry and confirmed spectroscopically
 appear to be true AGB C stars.
The task is more  difficult when one wishes to explore
regions at lower latitude, down to $|b|$\,$\sim$\,$20^{\circ}$, 
as we have attempted in this study, because contamination
by young  stellar  objects with similarly red colours becomes greater.
In the following, we describe new findings of C stars in the distant halo, and 
especially the discovery of a very distant C star at about 130\,kpc from the
Galactic Centre.

\section{Selection of candidates}

The selection method for  cool halo AGB C stars has been described in Paper~I.
Here, we briefly recall it and mention some changes that have been made
since Paper~I. Carbon star candidates are now selected from the
final, full-sky release of the 2MASS photometric 
catalogue\,\footnote{{\tt www.ipac.caltech.edu/2MASS}}.
Apart from a few exceptions, we impose limits $K_s < 13.5$, $J-K_s > 1.3$.
The limit in $K_s$ is sufficiently faint to allow detection of AGB C stars at 
large distances in the halo,
and the colour limit is imposed to avoid increasing contamination of
candidates by M type stars.
$J-H$ and $H-K_s$ colours are requested from 2MASS such that the representative point
in a $JHK_s$ colour-colour diagram lies within 0.15\,mag. of the relatively
narrow locus formed by the known  cool C  stars at high  galactic
latitude ($|b| > 30$$^{\circ}$) (see Paper I). This narrow  locus of C stars in the
$JHK_s$ diagram is seen well, for example, in Fig.\,2 of Nikolaev \& Weinberg 
(\cite{nikolaev00}).
Another change compared to Paper~I is that
we consider targets down to the  latitude $|b| = 20$$^{\circ}$,
giving especially attention  to candidates that  are faint in $K_s$, that is,
$K_s$  typically greater than $\sim$ 8.0. This is done with the idea that,
if the targets actually are AGB C stars, they are distant and far from the
galactic plane, and do not belong to the common disk N-type C star population.

After 2MASS candidates are identified with the above photometric criteria, they
are systematically checked in  various catalogues with Simbad (CDS). They are
also examined in POSS digitized surveys and when possible, on the objective
prism plates of Byurakan Observatory (Gigoyan et al. \cite{gigoyan01}). 
Candidates already known as C stars are excluded, except for specific reasons such
as measuring a missing radial velocity or improving  uncertain
cases observed previously. These various checks result in many candidates being
discarded because they appear to be  M-type giants, galaxies, QSOs,
young stellar objects, T Tauri stars, M-type or even L-type dwarfs
(e.g. Kendall et al. \cite{kendall03}). Some cases can also be eliminated as artefacts by
inspecting the POSS and 2MASS images.

%%%%%%%%%%%%%%%%%%%%%%  table des coordonees ===================================
\begin{table*}[!ht]
	\caption[]{List of observed faint cool  halo carbon stars. Coordinates
	$\alpha$  and $\delta$ (J2000) are given in the object names
	(2MASS Jhhmmss.ss$\pm$ddmmss.s). $l$, $b$ are in degrees. The magnitudes
	$B$ \& $R$  are from the USNO-A2.0 as given in the 2MASS catalogue,
	 from  the APM database (Irwin \cite{irwin00})
	or from the USNO-B1.0 (Monet et al. \cite{monet03}),
	with uncertainties of the order of 0.4 mag. The  $J H K_s$ values are
	from  the 2MASS all-sky point source catalogue, with uncertainties
	of the order of 0.02-0.03 mag.}
	\begin{flushleft}
	\begin{tabular}{lcrrrrrrrrrl}
	\noalign{\smallskip}
	\hline
	\hline
	\noalign{\smallskip}
No.&  2MASS name  & $l$~~~~ & $b$~~ & $B$ & $R$~~ & $B$-$R$ & $J$~~ & $H$~~ & $K_s$~~ & $J$-$K_s$&Note\\
	\noalign{\smallskip}
	\hline
 31& \object{2MASS J001655.77 $-$440040.6} &  323.07 & $-$71.74 & 17.6 &14.1 & 3.5 &  9.821 &  8.381 &  7.074 & 2.747&  new\\
 32& \object{2MASS J082915.12 +182307.2}   &  206.23 & +29.57 & 16.1 &11.5 & 4.6 &  8.748 &  7.709 &  7.067 &  1.681&   new\\
 33& \object{2MASS J082929.03 +104624.1}   &  214.18 & +26.61 & 19.3 &13.0 & 6.3 & 10.260 &  9.008 &  8.138 &  2.122&   new\\
 34& \object{2MASS J085418.70 $-$120054.0} &  239.10 & +20.47 & 19.0 &12.4 & 6.6 & 10.753 &  9.213 &  8.020 &  2.733 &  new\\
 35& \object{2MASS J085955.71 $-$775305.4} &  292.01 & $-$20.27 & 20.1 &14.3 & 5.8 & 12.865 & 11.557 & 10.554 &  2.311& new\\
 36& \object{2MASS J101037.00 $-$065113.9} &  248.12 & +38.35 & 16.9 &12.9 & 4.0 &  9.952 &  9.073 &  8.527 &  1.425& new\\
 37& \object{2MASS J114142.36 $-$334133.2} &  286.69 & +26.97 & 18.5 &16.3 & 2.0 & 13.848 & 12.802 & 12.178 &  1.670& Gizis\\
 38& \object{2MASS J121416.94 $-$090050.0} &  287.66 & +52.75 & 18.4 &15.3 & 3.1 & 13.222 & 12.355 & 11.875 &  1.347& new\\
 39& \object{2MASS J130953.60 +232541.0}   & 352.37 & +84.42  & 18.0 &14.1 & 3.9 & 11.751 & 10.464 &  9.486 &  2.265& Lieb.\\
 40& \object{2MASS J131901.11 +042021.2}   & 320.24 & +66.28  & 19.0 &15.7 & 3.3 & 13.074 & 12.245 & 11.847 &  1.227&new\\
 41& \object{2MASS J134723.04 $-$344723.3} & 315.77 & +26.68  & 17.7 &14.6 & 3.1 & 12.081 & 10.848 & 10.099 &  1.982&new\\
 42& \object{2MASS J143758.36 $-$041336.1} & 346.37 & +49.44  & 17.3 &14.5 & 2.8 & 12.265 & 11.411 & 10.903 &  1.362&new\\
 43& \object{2MASS J144631.08 $-$005500.2} & 352.23 & +50.60  & 19.2 &15.3 & 3.9 & 12.409 & 11.541 & 11.054 &  1.355& Marg.\\
 44& \object{2MASS J145726.98 +051603.4}   &   2.52 & +52.89  &  $-$  &18.2 & $-$  & 14.242 & 12.676 & 11.438 &  2.804&new\\
 45& \object{2MASS J152244.43 $-$123749.4} & 350.55 & +35.89  & 19.2 &15.5 & 3.7 & 13.080 & 11.983 & 11.433 &  1.647&new\\
 46& \object{2MASS J152723.59 +042827.8}   &   8.58 & +46.49  & 15.9 &12.0 & 3.9 & 10.159 &  9.020 &  8.127 &  2.032&new\\
 47& \object{2MASS J165227.70 $-$160738.4} &   3.82 & +17.31  & 19.7 &16.2 & 3.5 & 13.138 & 12.107 & 11.525 &  1.613& Gizis\\
 48& \object{2MASS J184650.29 $-$561402.8} & 339.65 & $-$21.66  & 18.9 &15.5 & 3.4 & 12.714 & 11.683 & 11.038 &  1.676&new\\
 49& \object{2MASS J191424.18 $-$782241.5} & 316.01 & $-$27.68  & 18.5 &14.7 & 3.8 &  9.662 &  8.486 &  7.434 &  2.228&new\\
 50& \object{2MASS J193138.52 $-$300230.5} &   9.09 & $-$21.27  & 19.5 &14.5 & 5.0 & 11.970 & 10.931 & 10.289 &  1.681&new\\
 51& \object{2MASS J193709.84 $-$353014.9} &  03.88 & $-$24.10  & 17.6 &15.0 & 2.6 & 11.950 & 10.787 & 10.222 &  1.728&new\\
 52& \object{2MASS J193734.14 $-$353237.6} &  03.86 & $-$24.20  &  $-$  &15.0 & $-$ & 11.266 & 10.001 &  9.125 &  2.141&new\\
 53& \object{2MASS J224350.35 $-$570123.3} & 331.15 & $-$52.54  & 14.7 &12.5 & 2.2 &  9.539 &  8.523 &  7.995 &  1.544&new\\
 54& \object{2MASS J224628.52 $-$272658.2} & +24.98 & $-$62.31  & 19.4 &17.0 & 2.4 & 14.867 & 13.890 & 13.274 &  1.593&new\\
 55& \object{2MASS J224738.01 $-$782717.2} & 310.32 & $-$36.83  & 16.7 &12.9 & 3.8 & 10.023 &  8.907 &  8.214 &  1.809&new\\
     \multicolumn{11}{c}{Three warm carbon stars: one from TI98, and two from Christlieb et al. survey}\\
 56& \object{2MASS J031424.48 +074443.99} & 172.87 & $-$40.87    & 17.7 &15.8 & 1.9 & 13.770 & 13.062 & 12.805 & 0.965& TI98\\
 57& \object{2MASS J211720.79 $-$055047.60} &  45.55 & $-$34.95  & 16.5 &15.4 & 1.8 & 12.472 & 11.786 & 11.615 & 0.857& Chris.\\
 58& \object{2MASS J211742.10 $-$065711.00} &  44.41 & $-$35.56  & 16.6 &14.8 & 1.9 & 12.673 & 12.000 & 11.808 & 0.865& Chris.\\
         \noalign{\smallskip}
   \hline
\end{tabular}
\end{flushleft}

{\small Notes: The last column (note) indicates whether the halo C star is a new 
discovery of this work  or whether its C rich nature was already established in the
literature (i.e. Gizis \cite{gizis02}, Liebert et al. \cite{liebert00}, Margon et al. 
\cite{margon02},  TI98 = Totten \& Irwin \cite{totten98},
Christlieb et al. \cite{christlieb01}).
}
\end{table*}
%---------------------------------------------------------------------------------------

\section{Observations}

% --------------------- figure avec un spectre illustratif -----------------------
   \begin{figure*}[!ht]
    \resizebox{\hsize}{!}{\rotatebox{-90}{\includegraphics{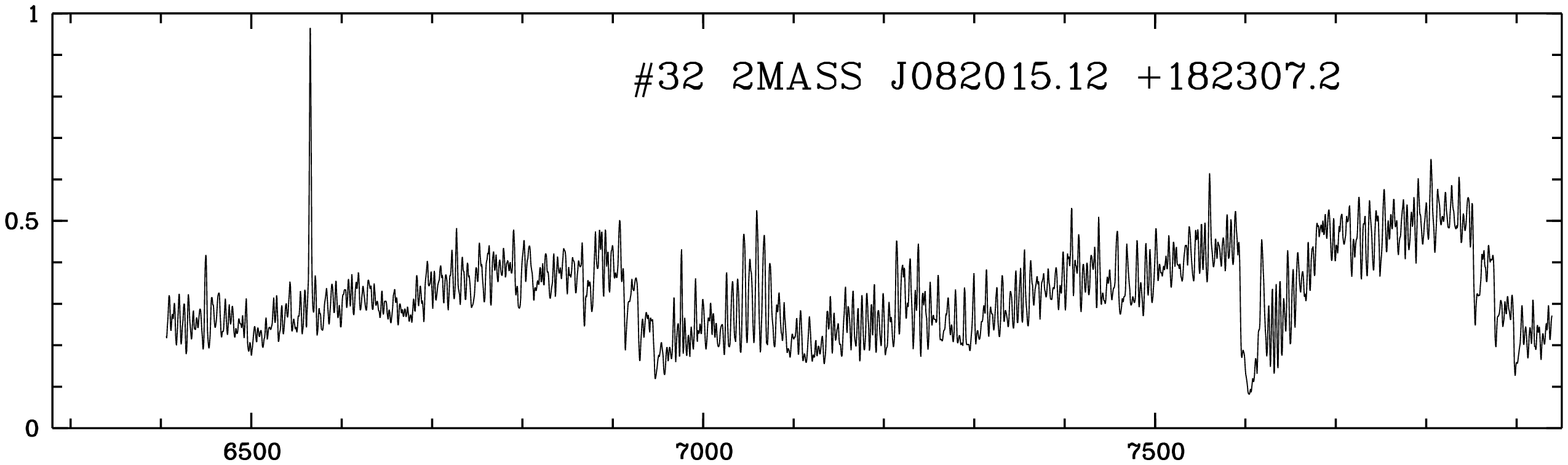}}}
     \caption[]{A typical spectrum, here for object \#32. The abscissa is
     wavelength in angstroms, and  the ordinate is flux in
     erg\,s$^{-1}$cm$^{-2}$\AA$^{-1}$, after division by
     an appropriate factor of 1.8 10$^{-13}$. The spectrum is dominated by CN and 
     C$_2$ features, and H$\alpha$ is in emission. The strong absorption band
     near 7600-7700\,\AA\ is telluric O$_{2}$. All other spectra are available in the
     electronic Appendix.}
     \label{fig01spec}
     \end{figure*}
%------------------------------------------------------------------------------------

Two runs were carried out at the NTT telescope of ESO (Chile) during
the nights  Apr 15--17, 2004 and  Aug 17--20, 2004. The instrument was EMMI used
in the medium dispersion mode (REMD) with a slit of 1$\arcsec$ and
 grating ``D/7''. This grating has
600 g\,mm$^{-1}$, is blazed at 6000\AA, and is used with an OG\,530 filter.
 The  detector was an MIT-LL 2048$\times$4096 CCD  with 15$\times$15 $\mu$m
 pixels used in  2$\times$2  binning and normal readout mode.
The resulting dispersion is 0.81\,\AA\, per pixel, and the resolution was
$\lambda$/$\delta\lambda$\,$\sim$\,2600 (corresponding to 115 km\,s$^{-1}$,  or
 $\delta\lambda$\,$=$\,2.7\,\AA\,at 7000\,\AA). The observed
wavelength range was 6406--7940\,\AA\, for the first run and
6312--7890\,\AA\,for the second run, the difference being due to an error
in instrumental setting. Each on-target exposure was followed immediately
(without depointing the telescope) by a spectral calibration exposure
made with a helium lamp (3\,sec) and an argon lamp (30\,sec). The exposure  times
on targets were from $\sim 10$ to $\sim$ 60 min.

Because the radial velocities of the targets are obtained by cross-correlation,
at least three  C rich stars (``templates'') with relatively well known radial
velocities were observed each night (see Paper~I and below). 
One photometric standard was also observed  to perform an approximate 
flux calibration.   Because the target objects are in general cool and variable; 
and since we ignore various
photometric effects, e.g. slit losses and sky transparency, the absolute flux
calibration can be only  indicative, but the slope of the spectra could  be
interesting to consider in future work.

 \section{General properties of the observed stars}

 A total of 54  selected candidates were observed  in the two ESO runs reported above.
 Among these 54 candidates were  included three that were already suspected to 
be C rich by Gigoyan et al. (\cite{gigoyan02}; \cite{gigoyan03}) but needed 
firm confirmation  (objects \#32, \#33, and \#46). From this sample, we obtained a total of 21
 confirmed new AGB C stars as listed in Table 2. In this table,
 the running number follows the numbers of the stars described in Paper~I.

 This list from \#31 to \#55 also includes 4 stars which were known  to be C rich
 with certainty but deserved more observations (objects \#37,\,\#39,\,\#43, and \#47).
  Objects \#37 and \#47 had been noted  C-type in Tables 1 and 2 of  Gizis (\cite{gizis02}), 
  who found them while searching for brown dwarfs in the TW Hya association, 
  but no spectra and radial velocity  were available
 for them. Note that  object \#47 has a low galactic latitude
 of $b = +17$$^{\circ}$, and therefore was not included in our initial list of 
 candidates, which is limited to $|b| > 20^{\circ}$.
 Object \#39 was discovered by Liebert et al. (\cite{liebert00}), while  \#43 was
 discovered in the Sloan survey by Margon et al. (\cite{margon02}); and  for both cases, 
 a measurement of their heliocentric radial velocity $v_{\rm helio}$ was desirable.

 Finally, in addition to this main sample, we included 
 three stars here that were observed in previous runs, but not reported
 in Paper~I. Of these 3 stars, \#56 is an object listed in TI98
 but without radial velocity, and the two others were selected from 
 the ESO-Hamburg sample of Christlieb et al. (\cite{christlieb01}). 
  All  three objects have $J-K_s$\,$\sim$\,$0.9$, which  allowed us 
 to compare the spectra of these warm objects
 to those having $J-K_s > 1.3$, which constitute our main sample.

  In order to estimate the efficiency of our survey for finding new C stars,
 we do not consider those 4 stars  that were
 already known to be C rich with certainty:
 2 from Gizis, 1 from Liebert, 1 from Sloan. There remain 54 candidates, 
 of which 21 were found to be
 new C stars. Our efficiency is therefore close to 39\%, somewhat
 better than  the  observation efficiency reported in  Paper~I (30\%).

 Concerning the global properties of the 25 stars numbered \#31 to \#55, one can note that
 the $R$-band magnitude has a range of 11.5 to 18.2, whereas the $B-R$ colour
 index is, in most cases, larger than 3. Only 5 cases have $B-R <3$. Two objects have
 no $B-R$\,: object \#44  has no $B$  available because of its extreme faintness, and
 its $J-K =2.8$ confirms that this object is very cool and possibly dusty.
 For \#52, no reliable $B$ and $R$ data  are available in catalogues
 because of  blending. We tentatively estimated its $R$ magnitude
 through a comparison  to star \#51, which  lies on the same digitized POSS2 red plate and is
 separated by only $\sim$ 5.5\,$\arcmin$ from \#52. A similar estimation was
 attempted for the $B$-band, but its quality was too poor to be considered.

 In the 2MASS data ($K_s$, $J-K_s$) of the same sample \#31 to \#55, one
 can note first that $K_s$ lies largely in the range 7.0 to 12.2, with a median of 10.1
 identical to the $K_s$-band median of the C stars discovered  in Paper~I. 
 In the list presented here, there is, however, a  remarkably faint object, \#54,  
 with $K_s$\,=\,13.27. This source
 will be discussed below in more detail.
 
 The colour index $J-K_s$ of our sample lies in the range 1.3 to 2.8, with one exception
 for object \#40 with  $J-K_s = 1.227$. This last object
 was found in a tentative selection where we sought
 candidates having $1.2$\,$<$\,$J-K_s$\,$<$\,$1.4$ with added  constraints
 $K_s > 11$ and  $B-R > 3$.  Thus, our sample (\#31 to \#55) is made of
 carbon stars that are red, with $J-K$ larger than 1.2. This indicates
 that these stars cannot be  early R-type stars, because early R-type objects
  have $J-K$ between 0.4 and 1.2, as can be derived from  Table~1 
 in Knapp et al. (\cite{knapp01}).

\section {Spectra}
For clarity, the complete atlas of all spectra is
given in the electronic appendix, but for illustration the spectrum of \#32, which is
typical,  is displayed here (Fig.\,\ref{fig01spec}).
 The spectra  of the 25 objects \#31 to \#55
are very similar to  those of Paper~I with rising flux distributions.
Of these 25 cases, 11 clearly show  H$\alpha$ in emission (like \#32), a proportion of
 $\sim$ 45\% of cases, exactly as the sample of  Paper~I.
The last 3 objects \#56, \#57, and \#58  are bluer in $J-K_s$ ($\approx 0.9$) and have a
noticeably flatter continuum. Two clearly display H$\alpha$ in absorption. 
 In all spectra, the strong band at 7600-7700 \AA\, is from telluric O$_{2}$.

\section{Radial velocities}
 The radial velocities were determined by cross-correlating our program stars
 with at least three ``template'' stars of roughly the same apparent brightness,
 observed the same night and with previously  determined velocities.
 These templates were chosen in Table 4 of TI98. A
 supplementary one, IRAS 17446-7809,
 was taken from the list of Loup et al. (\cite{loup93}). This star is also known as
 USNO-B1.0\,0118-1022651 and its $R$-band magnitude is $\sim 11-12$.
 Uncertainties on the velocities of these templates are indicated in Table\,2,
 as given by TI98 for APM objects or estimated using the CO profile of Nyman et al.
 (\cite{nyman91}) for  IRAS 17446-7809. Consistency between the velocities of
 these templates was checked to be very good ($<$\,10\,km\,s$^{-1}$)
 when correlating them one to one.

 Beside these templates, we also re-observed APM\,$0915-0327$, for which TI98
 had given +79\,km\,s$^{-1}$, whereas 
 Mauron et al. (\cite{mauron04}) had found +95\,km\,s$^{-1}$.
 With our current data, all independent  correlations with the other templates
 provide  $v_{\rm helio}$\,=\,+106\,km\,s$^{-1}$ ($\pm$\,10\,km\,s$^{-1}$).
 The spectra showed nothing peculiar, and we eventually rejected this star
 as a radial velocity template.

 {\it Erratum}: in Paper~I, an error was made for star \#15, alias
  2MASS J172825.76+700829.9. Its heliocentric radial velocity is $-158$ and
  not $+158$\,km\,s$^{-1}$.

%%%%%%%%%%%%%%%%%%%%%%%%%%%%%%%%%%%%%%%%%%%%%%%%%%%%%%%%%%%%%%%%%%%%%%%%%%%%%%%%%%%%%%%%%
\begin{table}[!t]
\caption[]{Adopted template carbon stars used for deriving
radial velocities by cross-correlation. Their
velocities, noted $v_{\rm helio}$, are in km\,s$^{-1}$.}

	\begin{center}
        \begin{tabular}{lrrl}
        \noalign{\smallskip}
        \hline
	\hline
        \noalign{\smallskip}
 Star &  $v_{\rm helio}$ &  Run\\
\noalign{\smallskip}
\hline
\noalign{\smallskip}

\object{APM 1019$-$1136}   &   +126 $\pm \,4$   & April\\
\object{APM 1406+0520}     &  $-$21 $\pm \,5$   & April\\
\object{APM 1519$-$0614}   &   +105 $\pm \,4$   & April\\
\object{IRAS 17446$-$7809} &   $-$5 $\pm \,5$    & August\\
\object{APM 2213$-$0017}   &  $-$44 $\pm \,3$    & August\\
\object{APM 0207$-$0211}   &  $-$140 $\pm \,5$   & August\\

\noalign{\smallskip}
\hline
\end{tabular}
\end{center}
\end{table}
%%%%%%%%%%%%%%%%%%%%%%%%%%%%%%%%%%%%%%%%%%%%%%%%%%%%%%%%%%%%%%%%%%%%%%%%%%%%%%%%%%%%%%

  \section {Proper motions}

  Information on proper motions from the USNO-B1.0 catalogue of 
  Monet et al. (\cite{monet03})
  is available for all the objects, except one: \#51 which is blended with neighbours.
  For 19 objects, the USNO-B1.0 proper motion is zero, while
  for 8 objects, a non-zero proper motion  is provided in USNO-B1.0 and listed in Table 3.
  One can see that these proper motions are very small in general, apart from the
  case of \#58. In Paper~I, we found the same situation with a few objects having
  very small but non-zero motions. The conclusion of those detailed investigations
  was that these very small proper motions  were suspect, and probably artefacts. 
  Essentially the same analysis could be done here. 
  
\begin{figure}
   %[!ht]
    \resizebox{\hsize}{!}{\rotatebox{-90}{\includegraphics{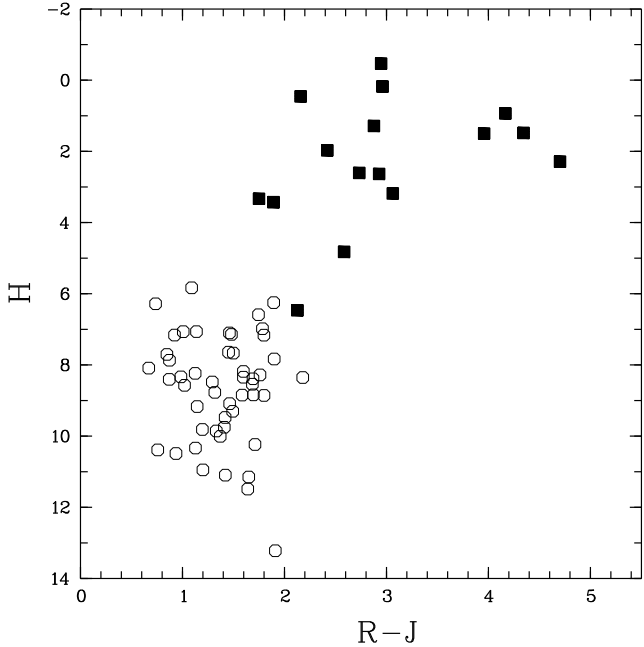}}}
     \caption[]{ The reduced proper motion $H = J + 5$\,log$_{10}(\mu)$ plotted
     as a function of the $R-J$ color index. Open circles are known dwarf
     carbon stars. Filled squares represent the 16 C stars of our sample
     for which USNO-B1 provides a non-zero  proper motion (see text).}
     \label{fig02mu}
     \end{figure}

   In addition, an analysis of  USNO-B1 data can be done by considering 
  the reduced proper motion  $H$\,$=$\,$J$\,$+$\,$5$\,log$_{10}(\mu)$. In Fig.\,\ref{fig02mu},
   we have plotted $H$ as a function of the color index $R-J$. The circles represent
  confirmed dwarf carbon stars with available 2MASS data  belonging  
  to the sample of Downes et al. (\cite{downes04}) or Lowrance et al. (\cite{lowrance03}). 
  The filled squares represent the 16 C stars that we found in Paper I and in 
  this paper, and for which the USNO-B1.0 proper motions are  not zero, but
  suspect. One can see that, with the exception of one object (\#58 at
  $R-J \approx 2.1$, $H \approx 6.5$), the locus of our C stars is very
  different from that of known dCs. This  shows that, even if
  the proper motions were correct, there would be no dwarfs  in our 
  sample of cool ($J-K \geq 1.2$) C stars.   
   
   We shall hereafter ignore these proper motions when they are small. 
   The case of \#58 for which a proper motion  of 57\,mas\,yr$^{-1}$ is
   relatively well measured will be treated in the next section.

% --------------------- figure avec un spectre illustratif -----------------------
%------------------------------------------------------------------------------

 % -----------------------------------------------------
  \begin{table}[!h]
          \caption[]{Data from the USNO-B1.0 database
  	for objects with non-zero proper motions.
  	 N is the number of detections on the scanned plates, and
  	Probab. is the total motion probability.}
  \begin{center}
          \begin{tabular}{crrrrr}
            \noalign{\smallskip}
          \hline
  	\hline
          \noalign{\smallskip}
   Object  & $\mu_{\alpha}$ cos $\delta$ & $\mu_{\delta}$  & N & Probab. \\
    \#     &(mas\,yr$^{-1}$)                     & (mas\,yr$^{-1}$)        & \\
  \noalign{\smallskip}
  \hline
  \noalign{\smallskip}
  36 & $+8\pm3$    & $+2\pm6$    &  5 &0.8 & \\
  44 & $-2\pm0$    & $-2\pm0$    &  3 &0.8 & \\
  45 & $+6\pm2$    & $+0\pm6$    &  4 &0.9 & \\
  47 & $-2\pm3$    & $+10\pm1$   &  4 &0.9 & \\
  53 & $-12\pm0$   & $+6\pm8 $   &  4 &0.9 & \\
  55 & $-8\pm15$   & $+16\pm2$   &  4 &0.9 & \\
  57 & $-4\pm5$    & $+10\pm2$   &  5 &0.8 &\\
  58 & $+50\pm13$  & $-28\pm10$  &  4 &0.9 &\\

  \noalign{\smallskip}
  \hline
  \end{tabular}
  \end{center}
  \end{table}
%--------------------------------------------------------------------------------

%%%%%%%%%%%%%%%%%%%%%%  table des proprietes ... et Notes %%%%%%%%%%%%
\begin{table*}
        \caption[]{Properties of the halo C stars. Quantities $l$, $b$, $R$, $K_s$, $J-K_s$ are
	repeated from Table 2 for information. The colour excess $E(B-V)$ is taken from
	 the maps of  Schlegel et al. (\cite{schlegel98}), in magnitudes, with uncertainty of the 
	 order of 0.01\,mag. The quantity $M_{K_s}$ is the  $K_s$-band absolute magnitude 
	 derived for each object, and is used to derive the distance to the Sun, $d$, in kpc.
	 Uncertainties (1$\sigma$) on $M_{K_s}$  and $d$  are $\sim 0.3$ mag. and 15\%, respectively.
	 $Z$ is the height above or below the galactic plane, in kpc. The measured 
	 heliocentric velocity $v_{helio}$ is in km\,s$^{-1}$, with uncertainty of the order 
	 of $\sim 10$ km\, s$^{-1}$ (1$\sigma$). The last column is the same as in Table 1.}
        \begin{flushleft}
        \begin{tabular}{lccrrrrrrrrl}
        \noalign{\smallskip}
        \hline
        \hline
        \noalign{\smallskip}
No.&  $l$ & $b$ & $R$ & $K_s$ & $J-K_s$ & $E(B-V)$ & $M_{K_s}$ & $d$ & $Z$ &$v_{helio}$ &Note\\
        \noalign{\smallskip}
        \hline
        \noalign{\smallskip}

 31 &  323.07  & $-$71.74 &14.1 &  7.074 &  2.747 & 0.007 & $-$7.25  &    7.5 & $-$7.0  & $-$38 &new\\
 32 &  206.23  & +29.57   &11.5 &  7.067 &  1.681 & 0.032 & $-$7.45  &    8.0 &    4.0  &+111 &new\\
 33 &  214.18  & +26.61   &13.0 &  8.138 &  2.122 & 0.035 & $-$7.55  &   13.5 &    6.0  & +90 &new\\
 34 &  239.10  & +20.47   &12.4 &  8.020 &  2.733 & 0.049 & $-$7.30  &   11.5 &    4.0  & +130 &new\\
 35 &  292.01  & $-$20.27 &14.3 & 10.554 &  2.311 & 0.192 & $-$7.45  &   39.0 & $-$13.5 & +270 &new\\
 36&   248.12  & +38.35   &12.9 &  8.527 &  1.425 & 0.035 & $-$7.10  &   13.0 &    8.0  & +306 &new\\
 37&   286.69  & +26.97   &16.3 & 12.178 &  1.670 & 0.071 & $-$7.40  &   82.0 &   37.0  & +144 &Gizis\\
 38&   287.66  & +52.75   &15.3 & 11.875 &  1.347 & 0.045 & $-$6.95  &   58.0 &   46.0  & +106 &new\\
 39&   352.37  & +84.42   &14.1 &  9.486 &  2.265 & 0.013 & $-$7.45  &   24.5 &   24.0  & +57 &Lieb.\\
 40&   320.24  & +66.28   &15.7 & 11.847 &  1.227 & 0.030 & $-$6.65  &   50.0 &   45.0  & +49 &new\\
 41&   315.77  & +26.68   &14.6 & 10.099 &  1.982 & 0.066 & $-$7.65  &   35.0 &   15.5  & +139 &new\\
 42&   346.37  & +49.44   &14.5 & 10.903 &  1.362 & 0.101 & $-$6.90  &   36.0 &   27.0  &+102 &new\\
 43&   352.23  & +50.60   &15.3 & 11.054 &  1.355 & 0.043 & $-$7.00  &   40.0 &   31.0  & +54 &Marg.\\
 44&     2.52  & +52.89   &18.2 & 11.438 &  2.804 & 0.040 & $-$7.25  &   55.0 &   44.0  & $-$13 &new\\
 45&   350.55  & +35.89   &15.5 & 11.433 &  1.647 & 0.162 & $-$7.35  &   55.0 &   32.0  & +126&new\\
 46&     8.58  & +46.49   &12.0 &  8.127 &  2.032 & 0.052 & $-$7.60  &   14.0 &   10.0  &+134 &new\\
 47&     3.82  & +17.31   &16.2 & 11.525 &  1.613 & 0.738 & $-$6.65  &   38.0 &   11.0  &+117 & Gizis\\
 48&   339.65  & $-$21.66 &15.5 & 11.038 &  1.676 & 0.083 & $-$7.40  &   49.0 &  $-$18.0 & +75 &new\\
 49&   316.01  & $-$27.68 &14.7 &  7.434 &  2.228 & 0.220 & $-$7.55  &    9.5 &  $-$4.5  &+177 &new\\
 50&     9.09  & $-$21.27 &14.5 & 10.289 &  1.681 & 0.121 & $-$7.40  &   34.0 &  $-$12.0 &+128 &new\\
 51&     03.88 & $-$24.10 &15.0 & 10.222 &  1.728 & 0.290 & $-$7.35  &   31.0 &  $-$13.0 &+137 &new\\
 52&     03.86 & $-$24.20 &15.0 &  9.125 &  2.141 & 0.288 & $-$7.60  &   21.0 &  $-$7.5  & +153 &new\\
 53&    331.15 & $-$52.54 &12.5 &  7.995 &  1.544 & 0.017 & $-$7.30  &   11.5 &  $-$9.0  &+131 &new\\
 54&     24.98 & $-$62.31 &17.0 & 13.274 &  1.593 & 0.022 & $-$7.35  &  130.0 &  $-$120 &+2  &new\\
 55&    310.32 & $-$36.83 &12.9 &  8.214 &  1.809 & 0.152 & $-$7.55  &   14.0 &  $-$8.5   & +378&new\\
            \multicolumn{7}{c}{Three warm carbon stars}\\
 56&  172.87   & $-$40.87 &15.8 & 12.805 & 0.965  & 0.317 & $-$2.0     &    8.5 &  $-$5.5   & $-$155&TI98\\
 57&   45.55   & $-$34.95 &15.4 & 11.615 & 0.857  & 0.180 & $-$2.0     &    5.0 &  $-$3.0   & $-$151&Chris.\\
 58&   44.41   & $-$35.56 &14.8 & 11.808 & 0.865  & 0.131 & $+$7.0     &    0.1 &  $-$0.06  & $-$70&Chris.\\

    \noalign{\smallskip}
   \hline
\end{tabular}
\end{flushleft}
\end{table*}
%%%%%%%%%%%%%%%%%%%%%%%%%%%%%%%%%%%%%%%%%%%%%%%%%%%%%%%%%%%%%%%%%%%%%%%%%%%%%%%%

%--------------------------------------------------------------------------------------
\section{Distances}

Determination of  distances was performed in the same way as in Paper~I and is based
on the $J-K_s$ and $K_s$ data. The $K_s$-band absolute
magnitude that was used  is in turn based on the $K_s$ magnitudes of C stars in the
Large Magellanic Cloud  (averaged in different bins in $J-K_s$)
$plus$ 0.5\,mag., where this supplementary term is due to the average difference between
C stars in the LMC and C stars in the Sgr dwarf galaxy (see Paper~I for more details).
The  effect of interstellar
reddening $E(B-V)$ is in general small  but was taken into account, because
several stars lie at at relatively low galactic latitude. The relations between $E(B-V)$ and
the extinction in $J$ and $K_s$ were taken from the 2MASS documentation,
i.e. its extinction calculator (Schlegel et al. \cite{schlegel98}). We adopted
$E(J-K)$\,$=$\,0.537\,$E(B-V)$ and $A$$_{\rm K}$\,=\,0.367\,$E(B-V)$, and
obtained the results listed in Table 2.

 The intrinsic $(J-K_s)_{\rm 0}$  of the 3 last warm stars of Table~1 can
 be obtained after correction of reddening, and one finds  $(J-K_s)_0$ between
 0.76 and 0.80.  This index suggests that they might be  {\it early} R-type giants, but
 certainly not {\it late} R-type giants. For instance, this $J-K_s$ value is bluer by at least
 0.3 mag. than all the $J-K_s$ values of the late R-type sample 
 of Knapp et al. (\cite{knapp01}),
 who found that the $K_s$-band absolute magnitude of early R-type giants
 is on the order of $-2.0 \pm 1$, and that the early R-type stars are clump giants.
 
 If we adopt the view that all three stars are clump giants, i.e. with 
  $M_{\rm K_s} \sim -2$, then their distances are found to be 
  8.6, 5.1, and 5.6\,kpc for \#56, \#57, and \#58, respectively. This, 
  however, cannot be true for \#58, because such a luminosity would imply
  an unfeasibly large  transverse velocity, $v_{\rm t}$. Since
  $v_{\rm t}$(km\,s$^{-1}$)\,=\,4.75\,$\mu$(mas\,yr$^{-1}$) $\times$  $d$(kpc),
  one would find  $v_{\rm t}$ $\sim$ 1500\,km\,s$^{-1}$, far too great
  for a star in our Galaxy. Therefore, significantly
  lower luminosity than that of a clump giant is necessarily implied, 
  and \#58 is a dwarf carbon (dC) star,
  provided the USNO-B1.0 data are correct. Its $K_{\rm s}$ absolute magnitude 
  might be on the order of $\sim +7$
  (Lowrance et al. \cite{lowrance03}). Then
  one finds $d$\,$\sim$\,90\,pc and $v_{\rm t}$\,$\sim$\,25\,km\,s$^{-1}$, consistent
  with the radial velocity of 70\,km\,s$^{-1}$. 
   An additional indication that
  \#58 is a dwarf is the presence of strong absorption NaI lines, 
  which are not seen in the spectra of the two other stars for which the NaI lines 
  were observed (\#56 and \#57).
%%%%%%%%%%%%%%%%%%%%%%%%%%%%%%%%%%%%%%%%%%%%%%%%%%%%%%%%%%%%%%%%%%%%%%%%%%%%%%%%%%%%%%%%%%%%%%%%

 \section{Analysis and discussion}

% --------------------- figure avec les C etudiees plottees en XYZ-----------------------
   \begin{figure*}[!t]
    \resizebox{\hsize}{!}{
    {\rotatebox{-90}{\includegraphics{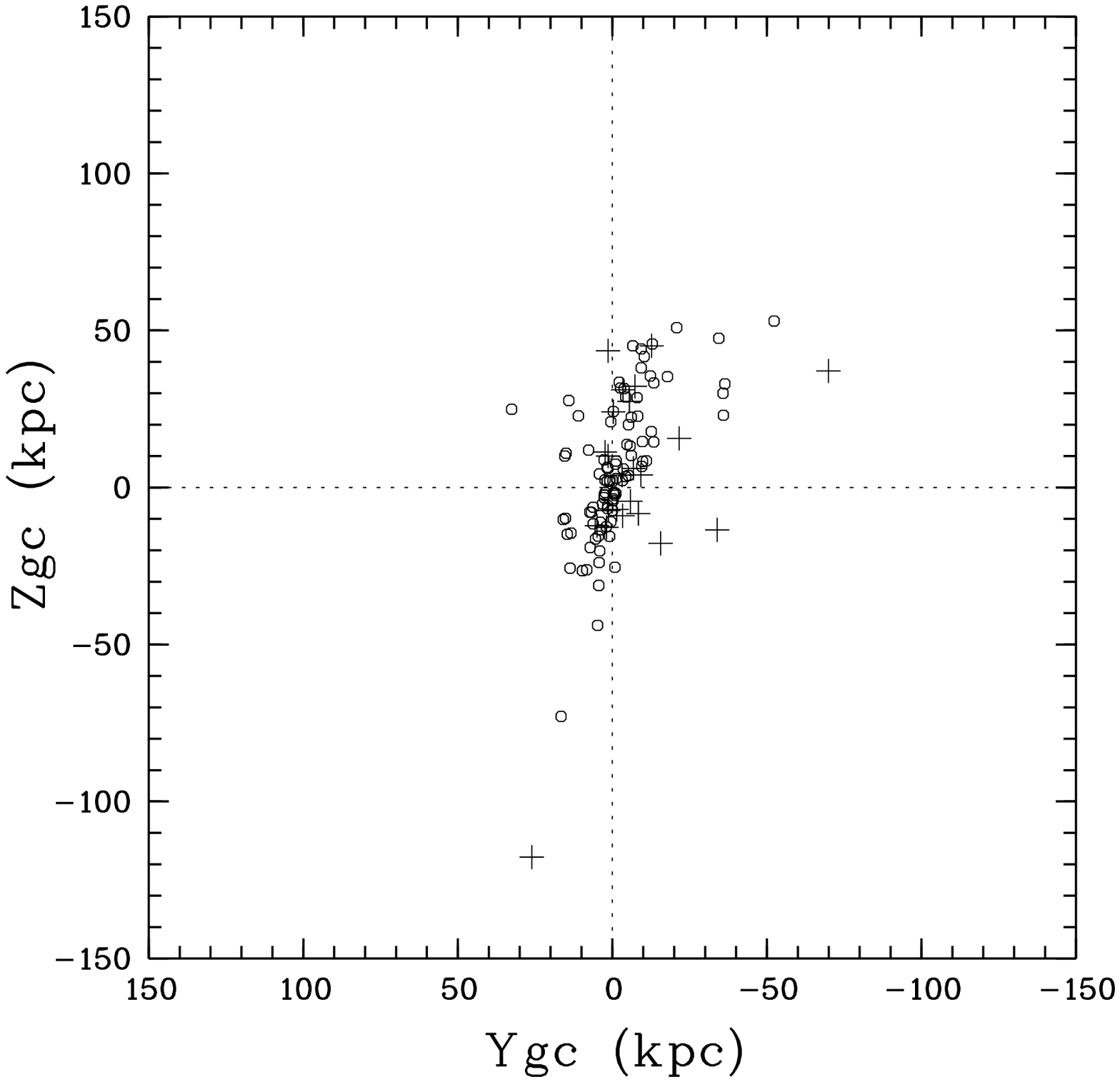}}}
    {\rotatebox{-90}{\includegraphics{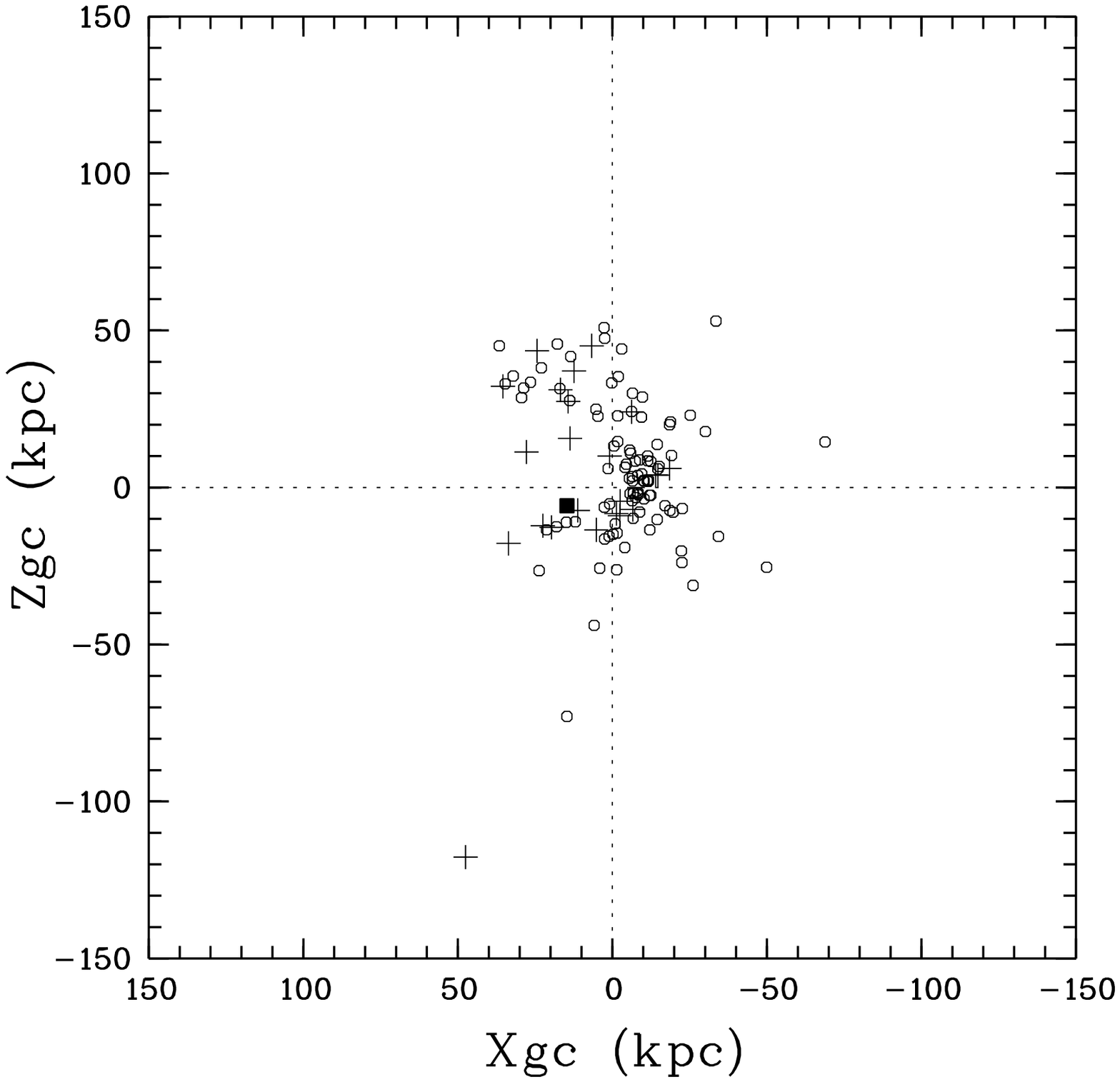}}} }

     \caption[]{ Plots in galactocentric $XYZ$ of the observed AGB C halo stars, with those found
     in this paper indicated by a cross. {\it Left panel}: the  C stars trace the Sgr Stream
     seen nearly edge-on, with a very distant C star to the South, and several C stars
     off the plane in the North. {\it Right panel}: the Sun is at
     $X$\,=\,$+8.5$\,kpc, $Z$=0 to the right of the origin, and
      the Sgr dwarf galaxy centre is indicated by a filled square at
     $X$\,=\,$14.7$\,kpc $Z$\,=\,$-5.8$\,kpc to the left of the origin.
     In the right panel, the scarcity of objects near the horizontal axis is due to
     the difficulty of finding very distant C stars close to the Galactic plane (see text).}
     \label{fig03csgr}
     \end{figure*}
%====================================================================================

% --------------------- figure avec les simulations de Law et al -----------------------
   \begin{figure*}[!t]
    \resizebox{\hsize}{!}{
    {\rotatebox{-90}{\includegraphics{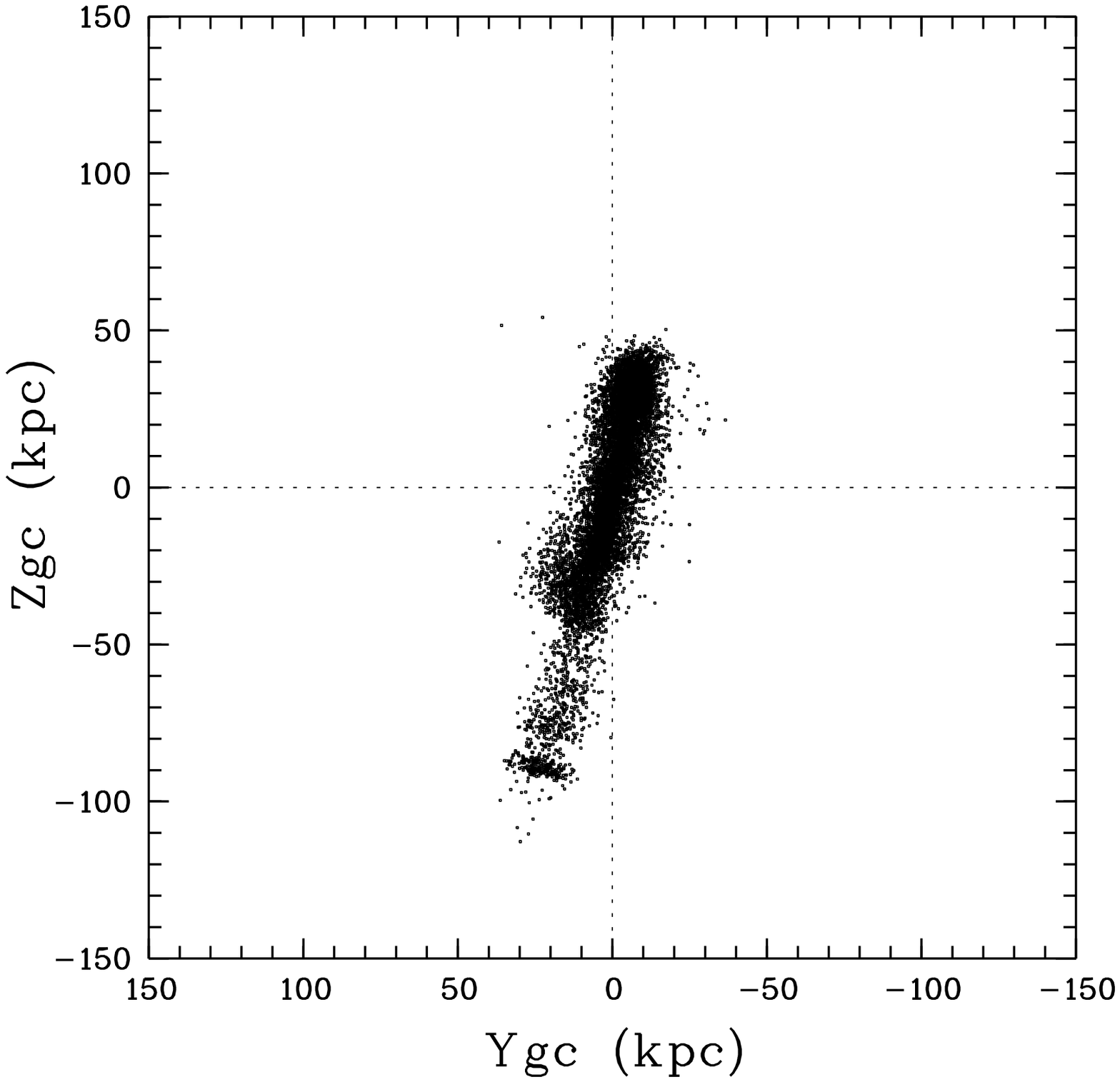}}}
    {\rotatebox{-90}{\includegraphics{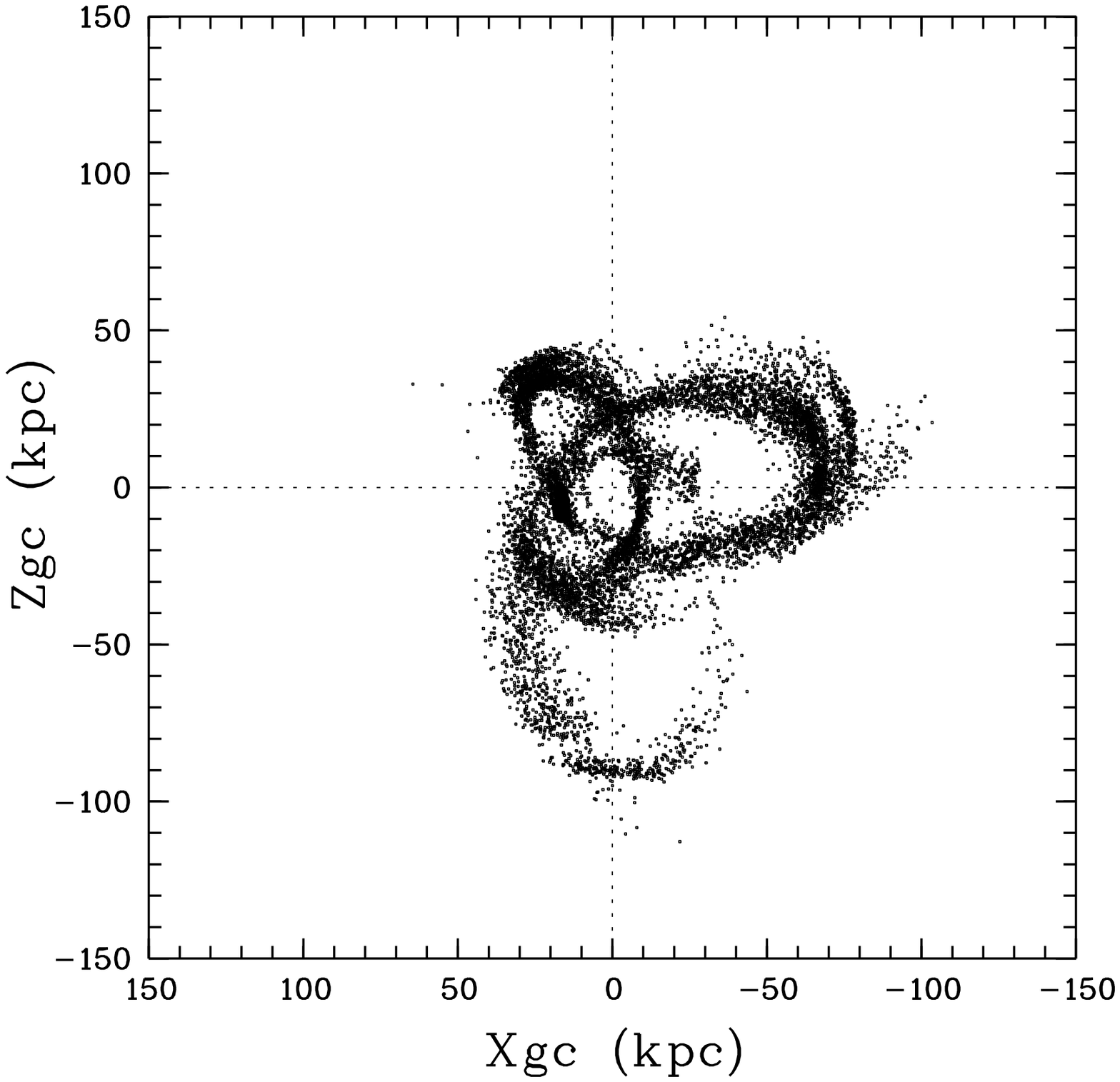}}} }
    \caption{Same plots in $XYZ$ of the model of the Sgr Stream, from the N-body simulations of 
    Law et al. (\cite{law04}). This simulation corresponds to the case of a spheroidal Galactic
    potential.}
    \label{fig04law}
    \end{figure*}
%======================================================================================    

 When the sample of cool C stars investigated above (\#31 to \#55) is added 
 to previous similar cases (Paper~I,
 TI98), one obtains a total of 119 stars; therefore, it is interesting to geometrically represent 
 this population of C stars in the halo. In the  two panels of Fig.\,\ref{fig03csgr}, we plotted 
 all known halo AGB C stars in the galactocentric system.  In this system,
 the Sun is at $X_{\tiny \sun}$\,=\,$-8.5$\,kpc, $Y_{\tiny \sun}$\,=\,$Z_{\tiny \sun}$\,=\,0.
 The $Z$ axis is toward the North galactic pole
 and  the $Y$ axis is towards the longitude $l$\,=\,$+90^{\circ}$. Note that the $X$ coordinate
 is the one chosen by Newberg et al.\,(\cite{newberg03}) or Bellazzini et al.\,(\cite{bellazini02}),
 but the opposite of that used in some other works. In our coordinates, the Sagittarius
 dwarf galaxy is at $X_{\tiny \rm Sgr}$\,=\,+14.6\,kpc, $Y_{\tiny \rm Sgr}$\,=\,+2.3\,kpc, and
 $Z_{\tiny \rm Sgr}$\,=\,$-$5.8\,kpc.   The C stars discovered in this paper are indicated
 by crosses, while those discovered in previous works are indicated by circles.
 Criteria to be included in these two  plots were that the objects have
 $J-K_s > 1.2$,  in order to select only AGB stars, and $|Z| > 1.5$\,kpc,  in order to
 eliminate  C stars which are almost certainly members of the galactic disk.
 
 For comparison to the data, in Fig.\,\ref{fig04law} we plotted the N-body simulation of the Sgr 
 Stream given by Law et al. 
 (\cite{law04})\,\footnote{{\tt http://www.astro.virginia.edu/$\sim$srm4n/Sgr/}}.
 We  chose to show the simulation for a spheroidal Galactic potential, since it is a
 compromise between the oblate and prolate cases, each of which have advantages and
 disadvantages, as discussed by Law et al. (\cite{law04}).  We emphasize that our goal 
 here is not to further improve this model, which already includes  as many as 11
 geometric and velocity  constraints based on detailed studies of the Sgr dwarf and its
 Stream (see Law et al., their Section 2). Rather, we intend to focus on the question of
 where the C stars are located and what this tells us, despite their scarcity and the 
 relative uncertainty of their distances on the order of $\pm 15$\% (1$\sigma$) 
 for a given $J$-$K_s$ (see Paper~I).

 In Fig.\,\ref{fig03csgr}, the left-hand diagram shows that the large majority of C stars 
 lie along a slightly inclined but roughly South/North straight line, which is 
 the plane of the Sgr Stream seen nearly edge-on
 (see, e.g., the upper panels of Fig.\,7  in Majewski et al. \cite{majewski03}), and this is also
 in fair agreement with the left-hand panel of Fig.\,\ref{fig04law}. The extension in $Z$ of the
 main cloud of points, ignoring the outliers, i.e. from $Z \sim -40$ to $Z \sim +45$\,kpc,  
 is similar in the observations and in the model, giving credence to the C star distance
 scale adopted here and in Paper~I.
 
  One particular
 object that we discovered is  \#54\,=\,2MASS\,J224628.52$-$272658.2. It is the 
 object plotted as a cross with the lowest $Z$ in Fig.\,\ref{fig03csgr}, left panel.  Its
 distance to the Sun, based on the use of its $K_s$ and $J-K_s$ values, is $d$ $\approx 130$\,kpc
 with $Z \approx -118$\,kpc, $Y \approx +26$\,kpc, and $X \approx +47$\,kpc.
 If our distance scale is correct, this star \#54 appears as the most distant {\it cool}  C star
 in the halo, more distant than any of the TI98 C stars. We note here that this
 star has a zero proper motion in the USNO-B1.0 catalogue, supporting the case for a {\it bona-fide} 
 distant AGB C star.
 
 The existence in the halo of similar distant objects is important.
 It is possible that other very distant {\it warm} C stars may be contained
 in the Sloan sample of Downes et al. (\cite{downes04}). More precisely, the Downes sample 
 contains 20 objects for which the proper motions are lower than 5 mas\,yr$^{-1}$, 
 that is, essentially undetected, and  for which  no classification as dwarf (D) 
 is given (they are marked U in their Table 1). The same authors also found 
 three certain members of the Draco dwarf galaxy, which is at a distance of 79\,kpc 
 (van den Bergh \cite{vandenbergh00}).
 The average of their Sloan $r$ magnitudes is $r \approx 17.0$, which provides an approximate 
 distance scale ($M_{\tiny r}$\,=\,$-2.5$). Of the 20 stars mentioned above,  
 9 have $r > 19$ and some of them, if not dwarfs, may be  more distant than 200\,kpc. 
 In summary, \#54 might not be the only case of a C star at more than 100\,kpc, 
 but attributing a definite giant type (rather than dwarf) to the stars of Downes 
 et al. is currently not possible.
 
   Star \#54 has another peculiarity: it  is
almost in the plane of the Sgr Stream. More precisely,
adopting the Newberg et al. (\cite{newberg03}) formulation and
 $XYZ$ coordinates described above, this plane is defined as:

\begin{equation}
    -0.064\,X + 0.970\,Y + 0.233\,Z = 0.232
\end{equation}

\noindent and it is readily found that object \#54 is only $5$\,kpc from this plane;
this distance to the plane is called $d_{\rm plane}$ hereafter.
 This shift is remarkably small given its large distance ($\sim$ 130\,kpc) 
 to the Galactic Centre.
 
 One can note in Fig.\,\ref{fig04law} that the Stream only  extends as far 
 as 100\,kpc to the Southern Galactic regions. Also, the agreement is rather poor in the 
 $XZ$ panel of Fig.\,\ref{fig04law}, that is, no Sgr debris particles are present close to \#54, with  
 about 40 kpc between \#54 and the closest points shown in Fig.\,\ref{fig04law}. This may suggest that
 we overestimate the distance of \#54 $provided$ we  choose 
 to trust the model points. Additional discoveries of  C stars, or RGB stars, with comparable
 distances, radial velocities, and galactic latitudes ($b < -60^{\circ}$) 
 would be most useful to clarify this point.

At the upper part of the left panel in Fig.\,\ref{fig03csgr}, one sees several stars which are off
the mean Sgr Stream plane mentioned above. Distances to the Stream plane can exceed
20\,kpc and are as large as 60\,kpc for one object with 
$X$\,$=$\,$+12$\,kpc, $Y$\,$=$\,$-70$\,kpc, $Z$\,$=$\,$+37$\,kpc. This
object is our \#37\,=\,\object{2MASS\,J114142.36$-$334133.2} 
(found by Gizis \cite{gizis02}). 
Its distance to the Sun is 82\,kpc, and this star has no sign of a proper motion in the
USNO-B1.0 catalogue. Its colours are typical of an AGB C star.

 There are 3 other objects for which the distance $d_{\rm plane}$ to the
Stream mean plane exceeds 30\,kpc. One is \#35 
from this paper  ($d = 39$\,kpc, $d_{\rm plane} = 36$\,kpc), and another
is \#4 from Paper~I ($d = 78$\,kpc, $d_{\rm plane} = 36$\,kpc). The last one
is a faint high latitude variable star J1714.9+4210 studied by Meusinger \& Brunzendorf
(\cite{meusinger01}), also mentioned in TI98 and 
Kurtanitze \& Nikolashvili (\cite{kurtanidze00}).
It is also \object{2MASS\,J171457.52+421023.4}. Meusinger \& Brunzendorf 
(\cite{meusinger01}) found that
this object is variable and  has a C rich spectrum. We verified that
its 2MASS colours are typical of a C star, located exactly on the narrow locus of the
AGB carbon stars in the $JHK_s$ diagram. Its has $K_s=11.17$ and 
$J-K_s = 1.354$, $d = 43$\,kpc, and 
$d_{\rm plane} = 37$\,kpc. None of these four stars have any measurable proper motion.
In summary, we have found 4 stars with typical AGB characteristics that are located
between 30 and 60\,kpc from the plane of the Stream. In addition, 
if we request  $|d_{\rm plane}| > 20$\,kpc, 
we get 9 stars, with only 1 on the left side in Fig.\,\ref{fig03csgr}, the left panel 
(the Meusinger \& Brunzendorf object), and 8 on the right side. This asymmetry is
clearly visible in Fig.\,\ref{fig03csgr} (left panel) and does not seem to be predicted by the model.

The right panel of Fig.\,\ref{fig03csgr} is very different from what the model shows in the right panel
of Fig.\,\ref{fig04law}. Observations show in $XZ$ that C stars are scattered within about 
50\,kpc from  the galactic centre with a few exceptions beyond this limit, while the model
displays much larger extensions to the right and to the South of the centre. However,
one has to take a number of observational limitations into account. In Fig.\,\ref{fig03csgr}, right panel,
the clump to the right of the centre is due to stars close to the Sun, and 
their numbers reflect the fact that they
were more easily discovered than those located far away from the Sun.
Another fact to take into account in
interpreting the data is a lack of points near the $Z=0$ plane. This
directly reflects the presence of the galactic plane, which surveys for halo
C stars naturally avoid. As a consequence, very few confirmed C stars can be seen
near the core of the Sgr dwarf Galaxy, to the left of the figure centre, indicated with coordinates
at $X_{\tiny \rm Sgr}$\,=\,+14.6\,kpc and  $Z_{\tiny \rm Sgr}$\,=\,$-$5.8\,kpc.

Despite these difficulties in interpreting the $XZ$ panel,
we note  an encouraging point of agreement: the observations indicate
many C stars in the region delineated by $X$ between 0 to +35\,kpc and
$Z$ between $+30$ and +50\,kpc. This region is $\sim$ 40\,kpc
above the Sgr dwarf centre (i.e. toward the North Galactic pole from Sgr)
and the numerical models show that this region is abundant in the
leading tidal debris of the Stream (see also
 Figs.~1 and 13 of  Law et al. \cite{law04}).

% ------------------------------------------------------------------------------- .
\section{Concluding remarks}

From this second paper on the  rare cool AGB C stars located in the Galactic halo,
the main results are the following. We  first discovered and/or documented 
25 new cool C stars yielded by colour selection from the 2MASS photometric catalogue, 
and included new data for a few serendipitous discoveries previously reported in the literature.
These stars  have a distance $|Z|$ away from the galactic plane larger than
$\sim$\,3\,kpc and up to $\sim$\,130\,kpc and cannot be 
ordinary N-type stars of the galactic disk. 

We then considered the totality of all AGB C stars known in the halo
(119 objects), and focused on their
locations in the  galactocentric $XYZ$ system. Then, we compared them with one of the
available N-body simulations of the Stream of the Sagittarius dwarf galaxy.  

The majority of these cool C stars originate in the tidal disruption of  the Sgr dwarf. 
A few remarkable objects deserve further study. For example, we discovered 
a very distant  ($\sim$ 130\,kpc) southern C star, which is almost exactly 
in the mean plane of the Sgr Stream and more distant than any similar object hitherto
documented.  
We also found four other remarkable C stars that are located off this plane, by large shifts, 
from 30\,kpc to 60\,kpc. None of these 4 objects seem to be dwarfs when 2MASS colours,
spectra, proper motions, and other properties are considered. For a  distance off the 
plane larger than 20 kpc, 9 stars were found. 

Because cool AGB C stars are rare, 
observations of much more numerous (but fainter) red giants lying on the RGB
are certainly needed to make our findings concerning the Sgr Stream
geometry more robust. In some sense, the C stars can be considered as the reddest and most 
luminous tracers of a debris which remains to be found with  deeper exploration.
In the future, it will  be interesting to
see whether a systematic spectroscopic survey of our fainter candidates($R$\,$\sim$\,17--21, 
$\sim$ 80 objects)  would result in the discovery of more of these AGB C stars very far 
away in the halo, i.e. at distances  from $100$ to $\sim$\,200\,kpc.

%=====================================================================

 \appendix

  \section{Spectra}

     \begin{figure*}

    \resizebox{\hsize}{!}{\rotatebox{-90}{\includegraphics{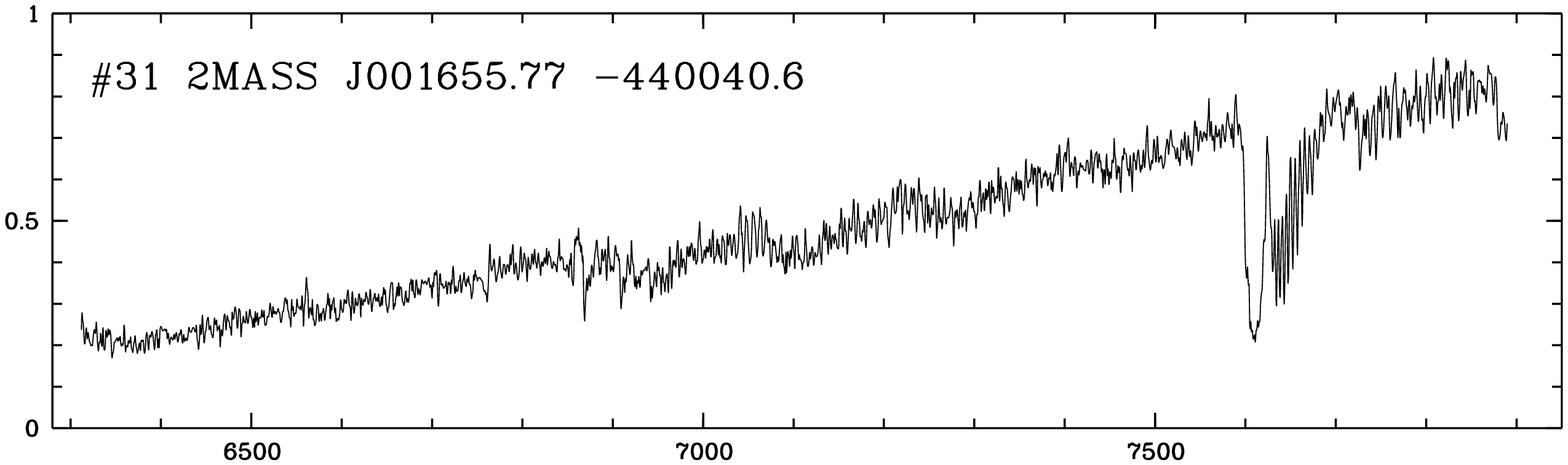}}}
    \resizebox{\hsize}{!}{\rotatebox{-90}{\includegraphics{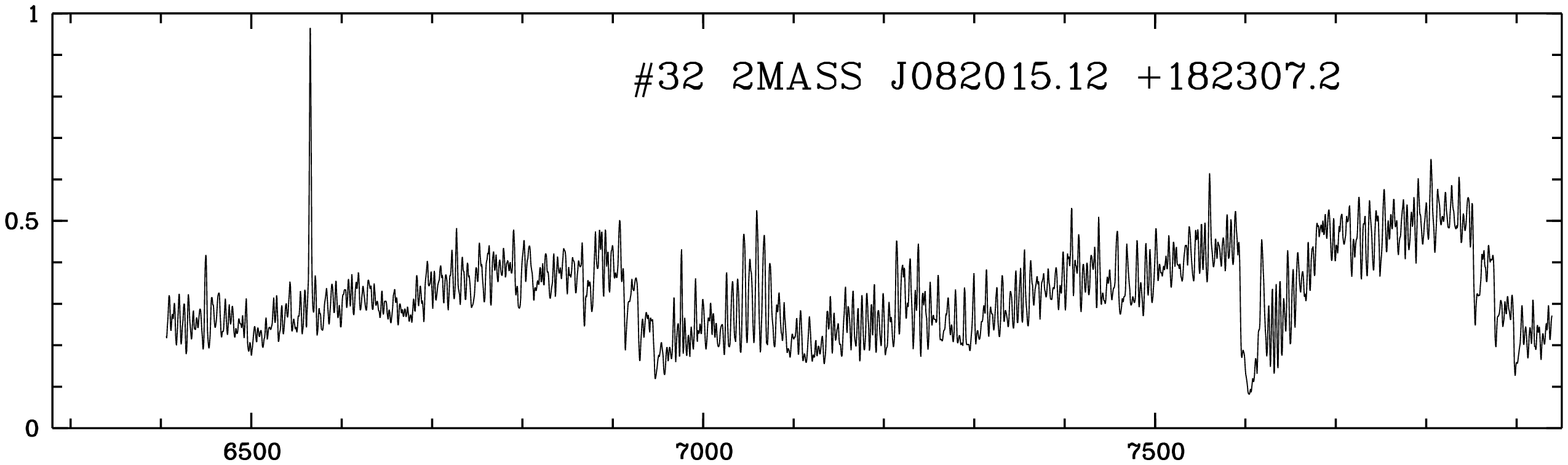}}}
    \resizebox{\hsize}{!}{\rotatebox{-90}{\includegraphics{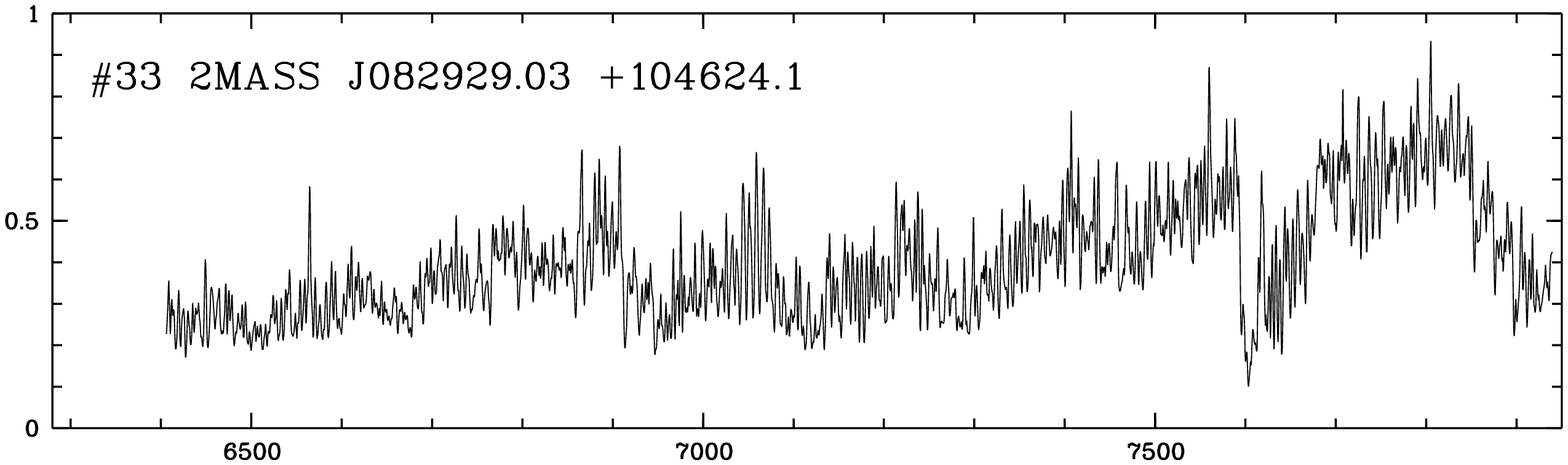}}}
    \resizebox{\hsize}{!}{\rotatebox{-90}{\includegraphics{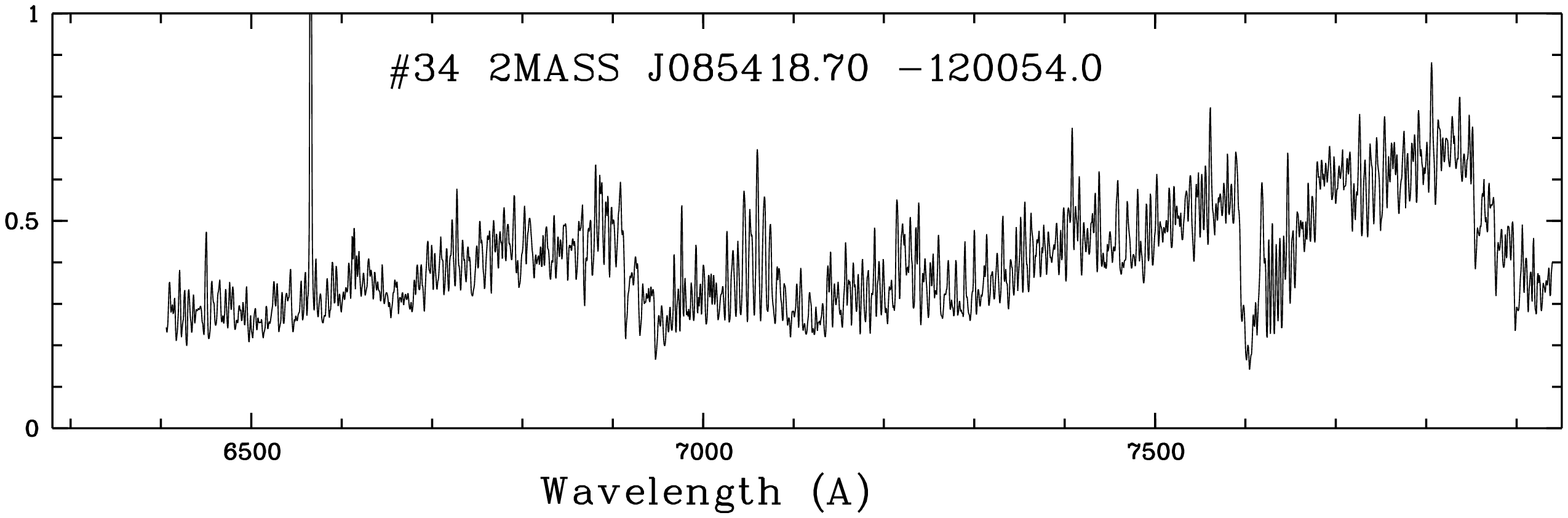}}}

       \caption[]{Spectra of objects 31, 32, 33, \& 34. In all these graphs,
        fluxes in ordinates are in erg\,s$^{-1}$cm$^{-2}$\AA$^{-1}$. Fluxes were divided by factors
        0.7 10$^{-14}$, 1.8 10$^{-13}$,  0.5 10$^{-13}$, \& 0.25 10$^{-13}$
        for objects 31, 32, 33, \& 34, respectively. }
   \label{spe1}
   \end{figure*}

     \begin{figure*}

    \resizebox{\hsize}{!}{\rotatebox{-90}{\includegraphics{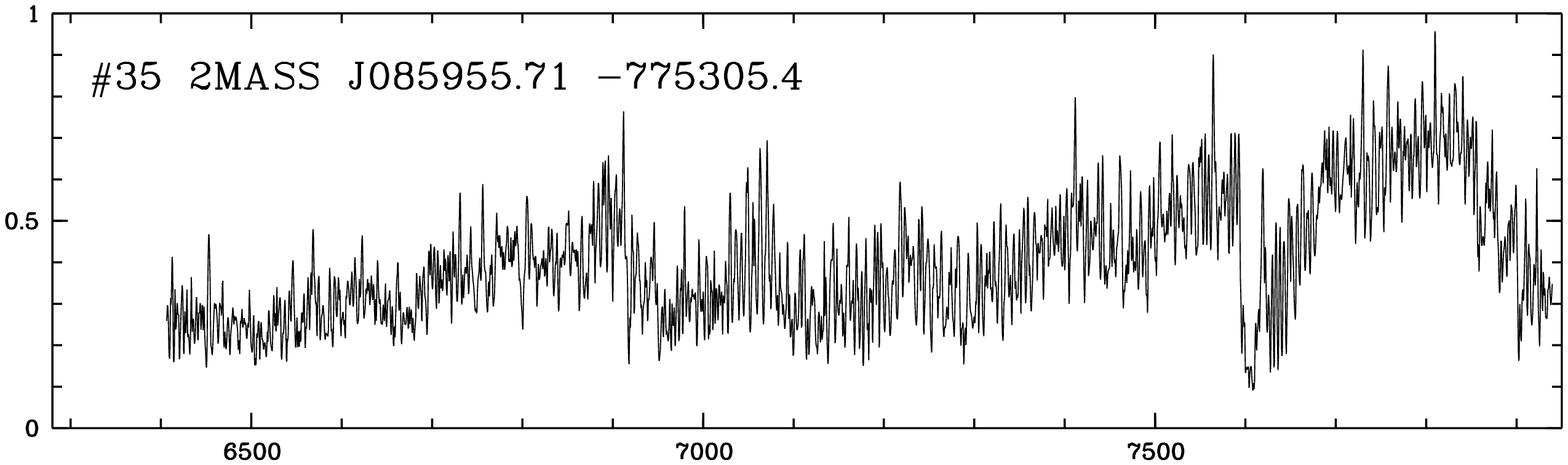}}}
    \resizebox{\hsize}{!}{\rotatebox{-90}{\includegraphics{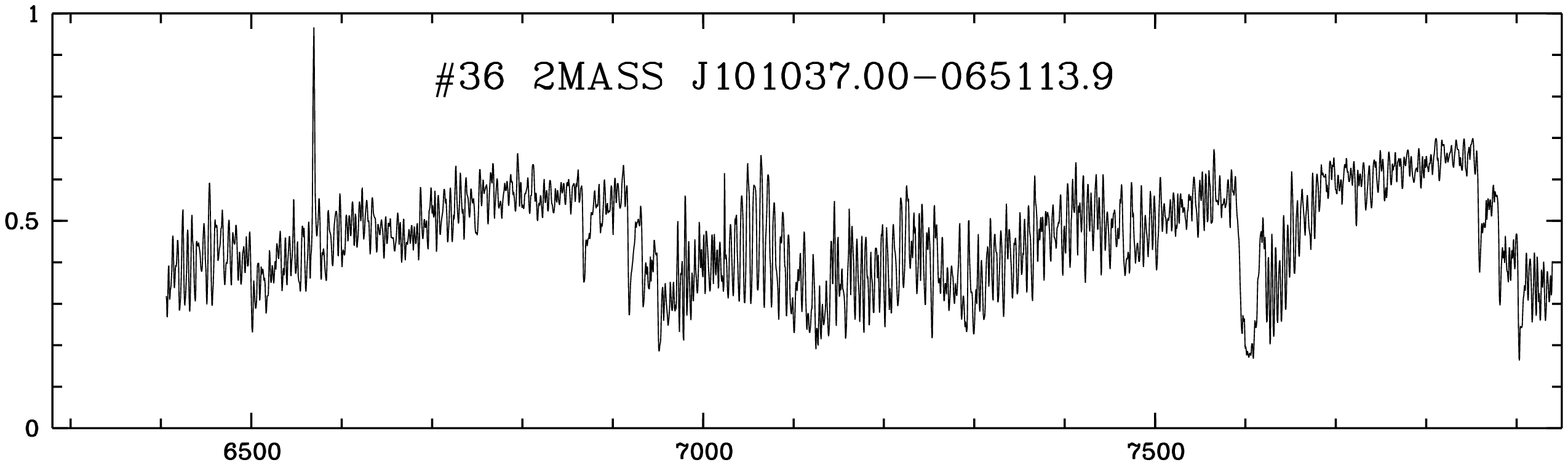}}}
    \resizebox{\hsize}{!}{\rotatebox{-90}{\includegraphics{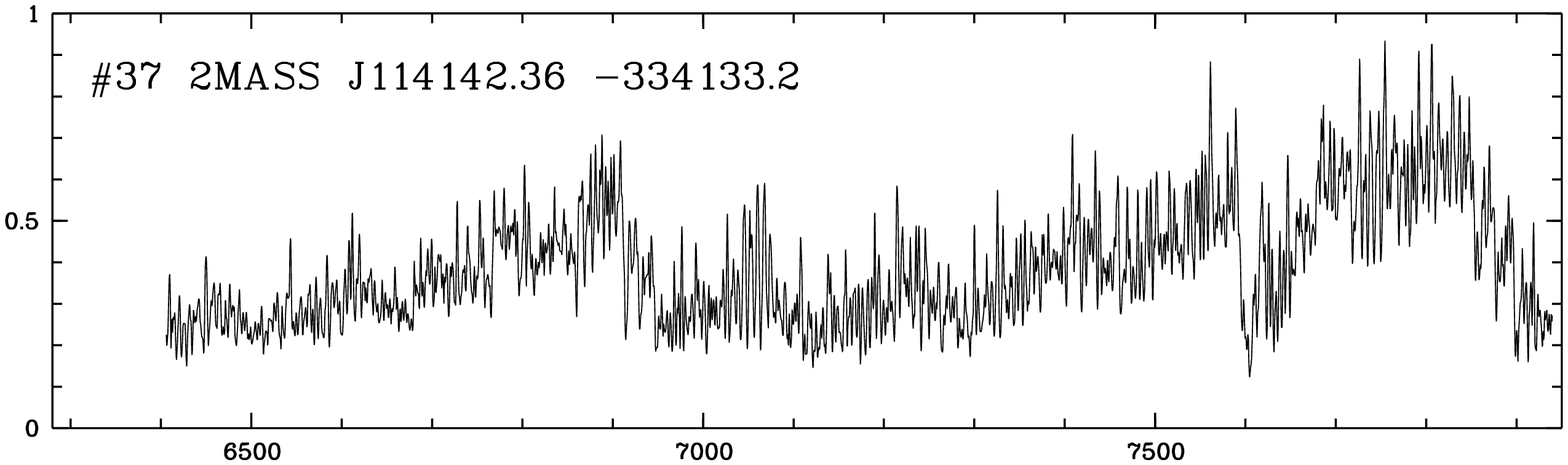}}}
    \resizebox{\hsize}{!}{\rotatebox{-90}{\includegraphics{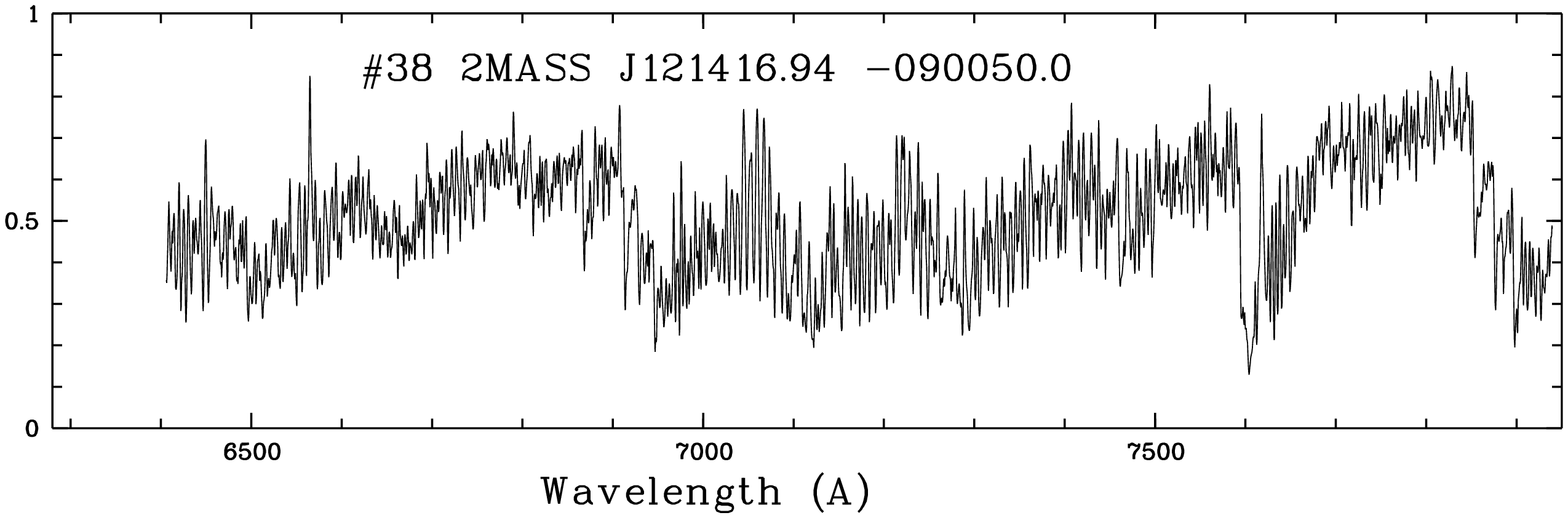}}}

       \caption[]{Spectra of objects 35, 36, 37, \& 38. In all these graphs,
        fluxes in ordinates are in erg\,s$^{-1}$cm$^{-2}$\AA$^{-1}$. Fluxes were divided by factors
        0.18 10$^{-14}$, 5.5 10$^{-15}$,  0.22 10$^{-14}$, \& 0.25 10$^{-14}$
        for objects 35, 36, 37, \& 38, respectively. }
   \label{spe2}
   \end{figure*}

 %%==============================================================
     \begin{figure*}

    \resizebox{\hsize}{!}{\rotatebox{-90}{\includegraphics{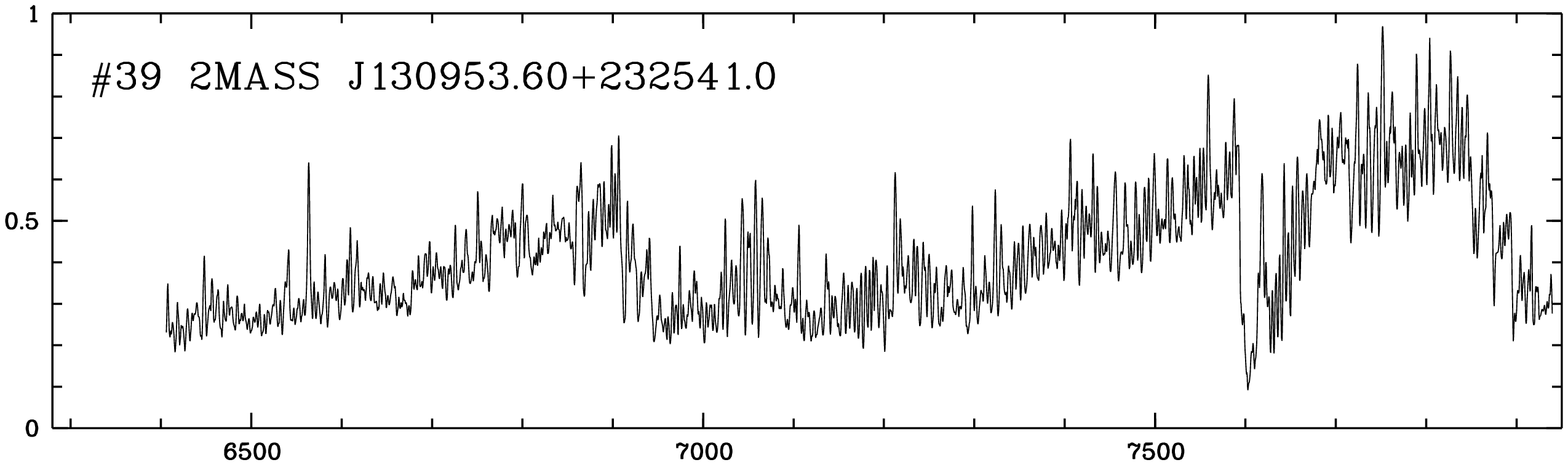}}}
    \resizebox{\hsize}{!}{\rotatebox{-90}{\includegraphics{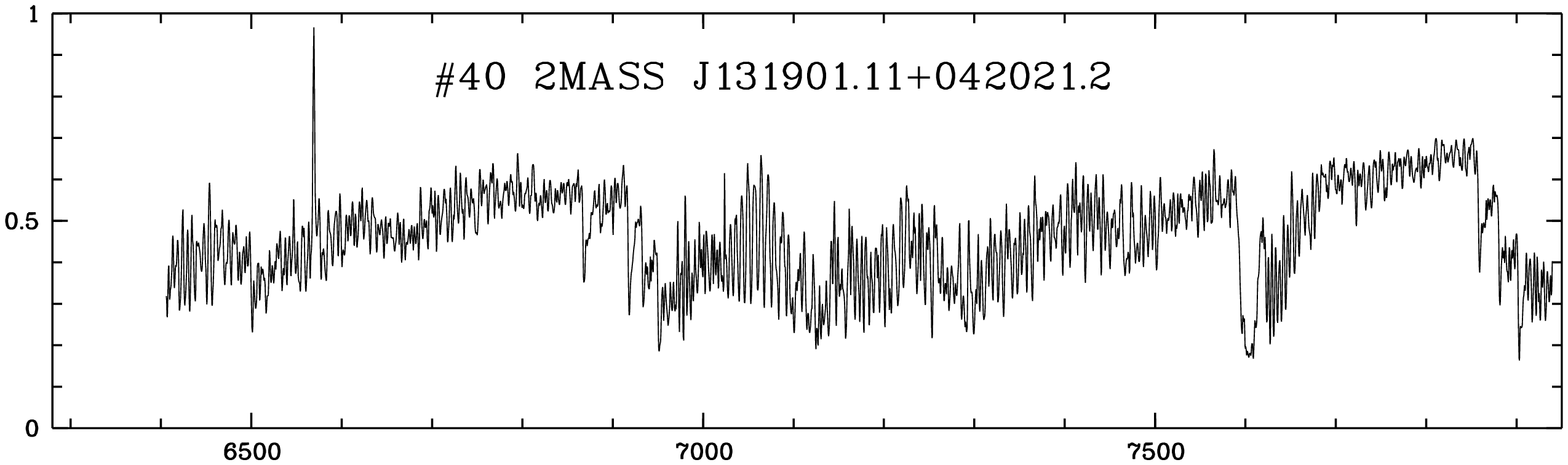}}}
    \resizebox{\hsize}{!}{\rotatebox{-90}{\includegraphics{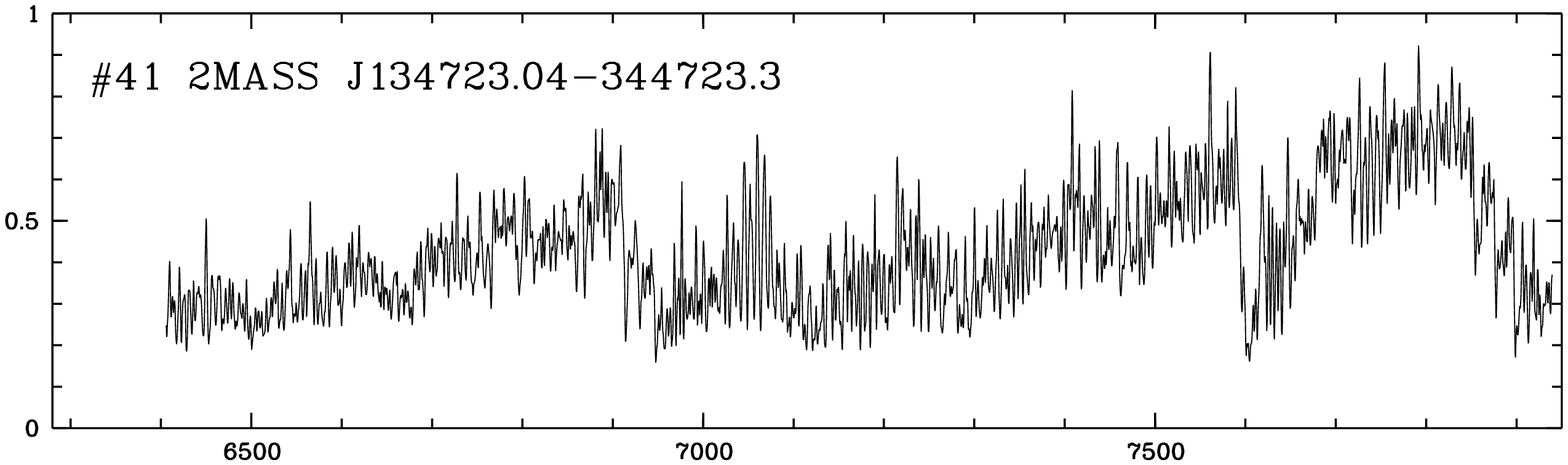}}}
    \resizebox{\hsize}{!}{\rotatebox{-90}{\includegraphics{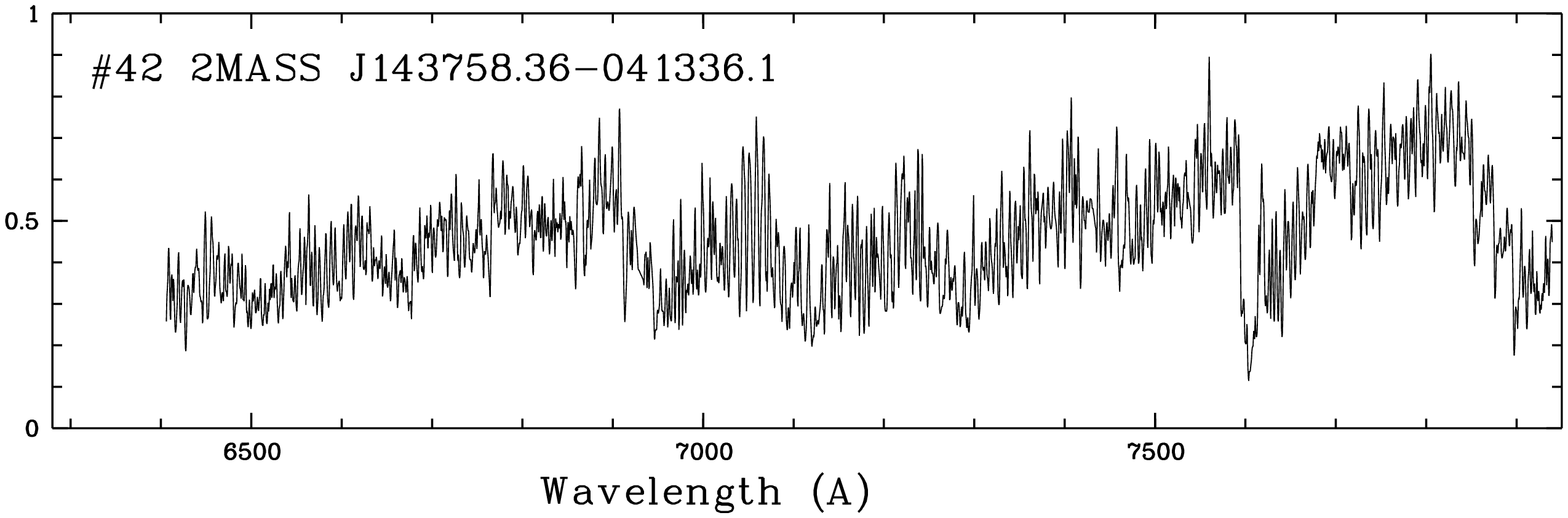}}}

       \caption[]{Spectra of objects 39, 40, 41, \& 42. In all these graphs,
        fluxes in ordinates are in erg\,s$^{-1}$cm$^{-2}$\AA$^{-1}$. Fluxes were divided by factors
        0.95 10$^{-14}$, 5.5 10$^{-14}$,  0.9 10$^{-14}$, \& 0.4 10$^{-14}$
        for objects 39, 40, 41, \& 42, respectively. }
    \label{spe3}
    \end{figure*}

%+==============================================
     \begin{figure*}

     \resizebox{\hsize}{!}{\rotatebox{-90}{\includegraphics{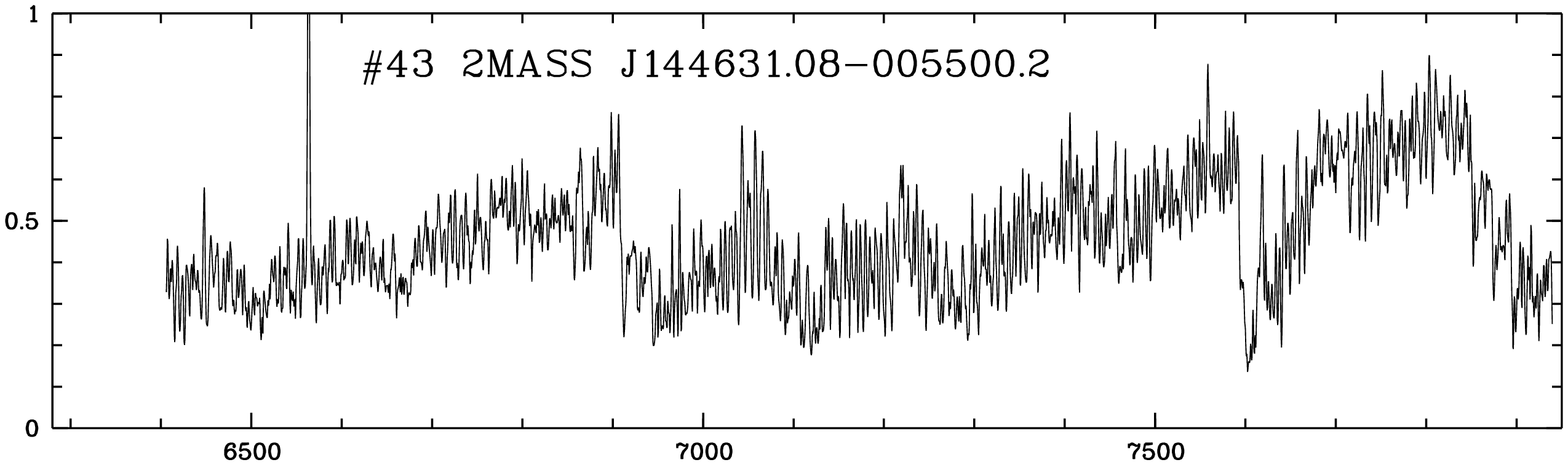}}}
     \resizebox{\hsize}{!}{\rotatebox{-90}{\includegraphics{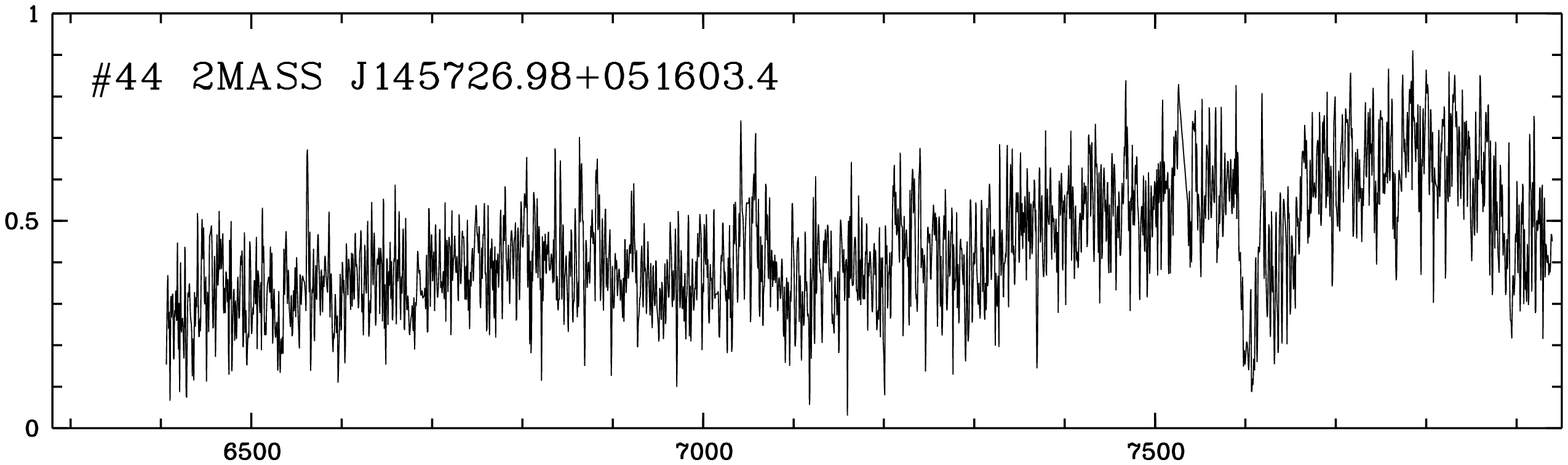}}}
     \resizebox{\hsize}{!}{\rotatebox{-90}{\includegraphics{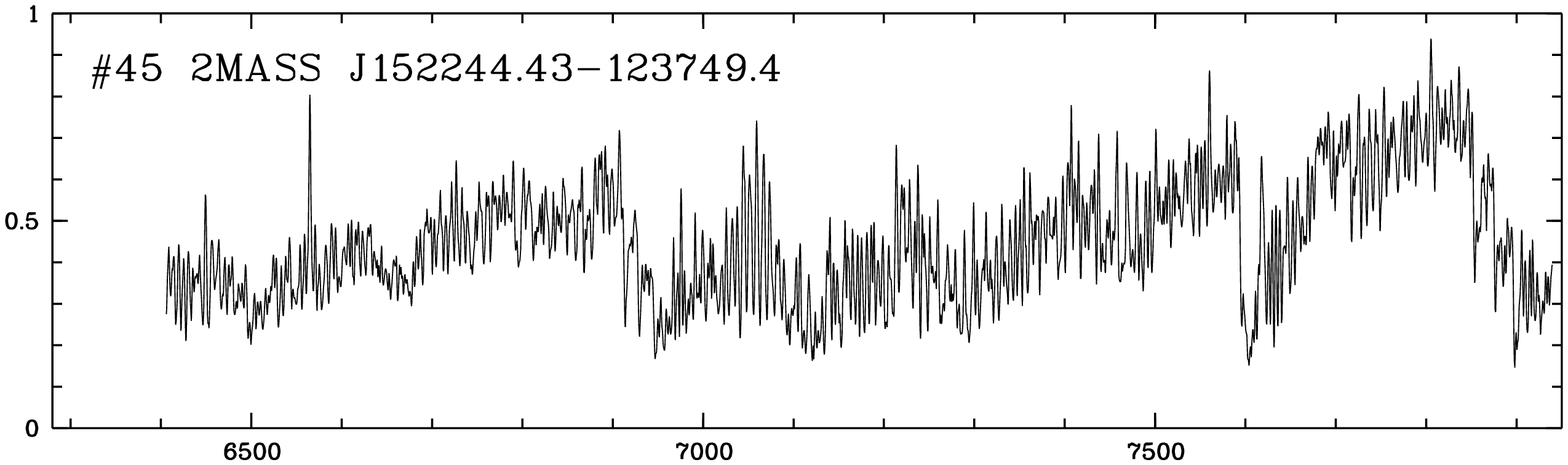}}}
     \resizebox{\hsize}{!}{\rotatebox{-90}{\includegraphics{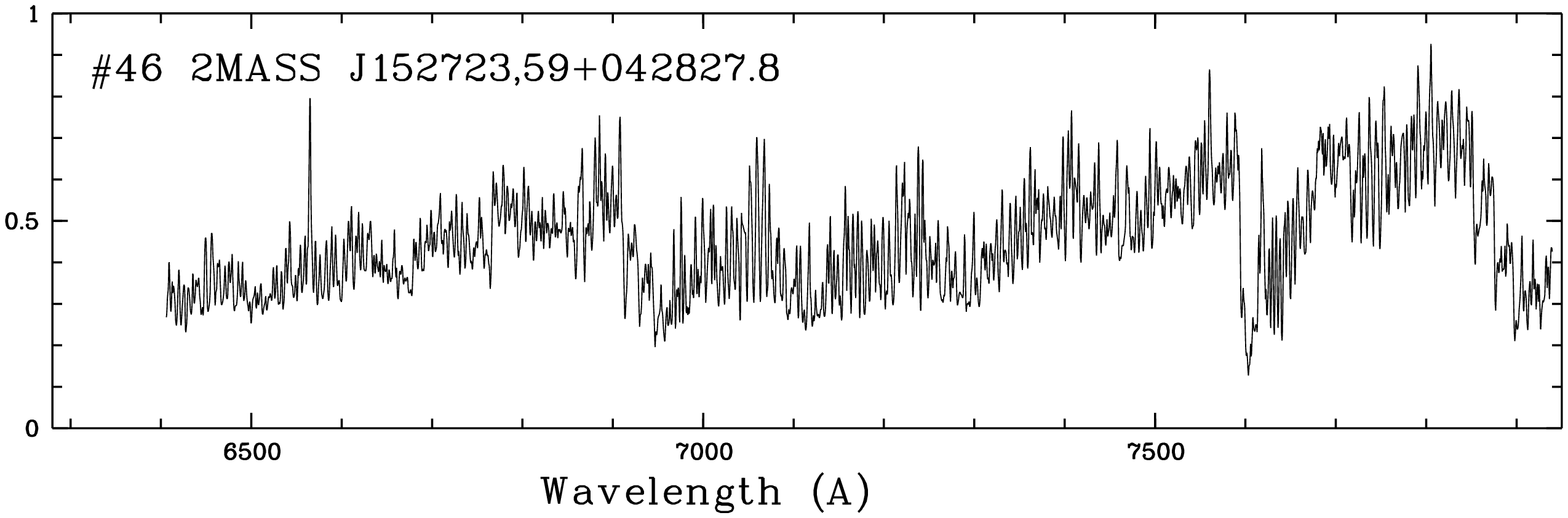}}}

      \caption[]{Spectra of objects 43, 44, 45, \& 46. In all these graphs,
       fluxes in ordinates are in erg\,s$^{-1}$cm$^{-2}$\AA$^{-1}$. Fluxes were divided by factors
       0.22 10$^{-14}$, 0.017 10$^{-14}$,  4.5 10$^{-15}$, \& 3.5 10$^{-14}$
       for objects 43, 44, 45, \& 46, respectively. }
  \label{spe4}
  \end{figure*}
 %===============================================================
   \begin{figure*}

   \resizebox{\hsize}{!}{\rotatebox{-90}{\includegraphics{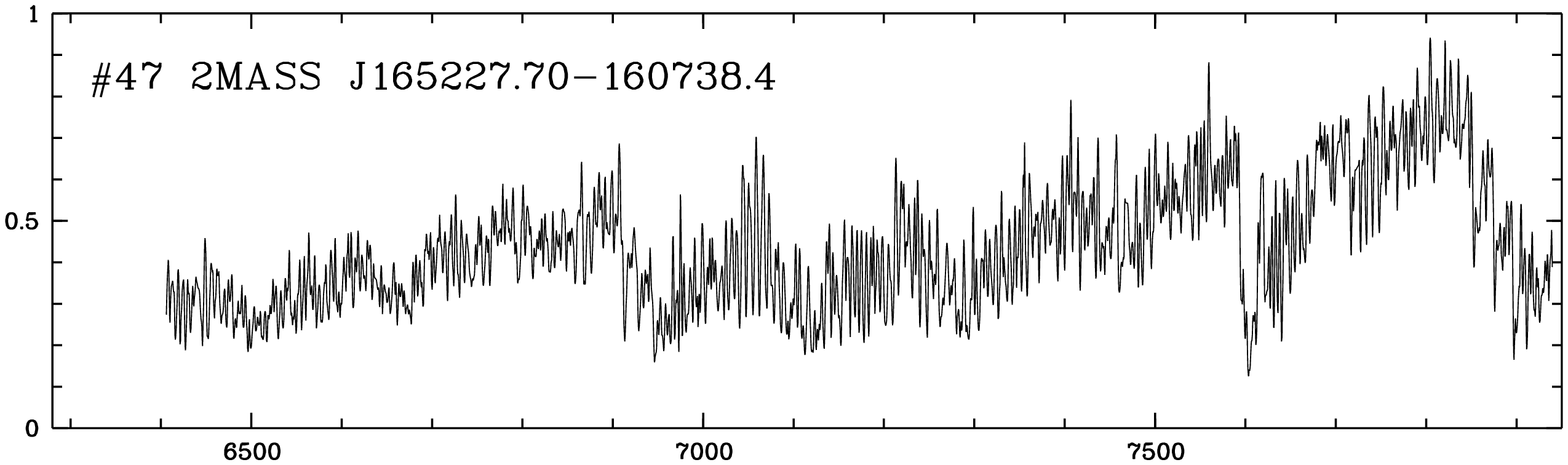}}}
   \resizebox{\hsize}{!}{\rotatebox{-90}{\includegraphics{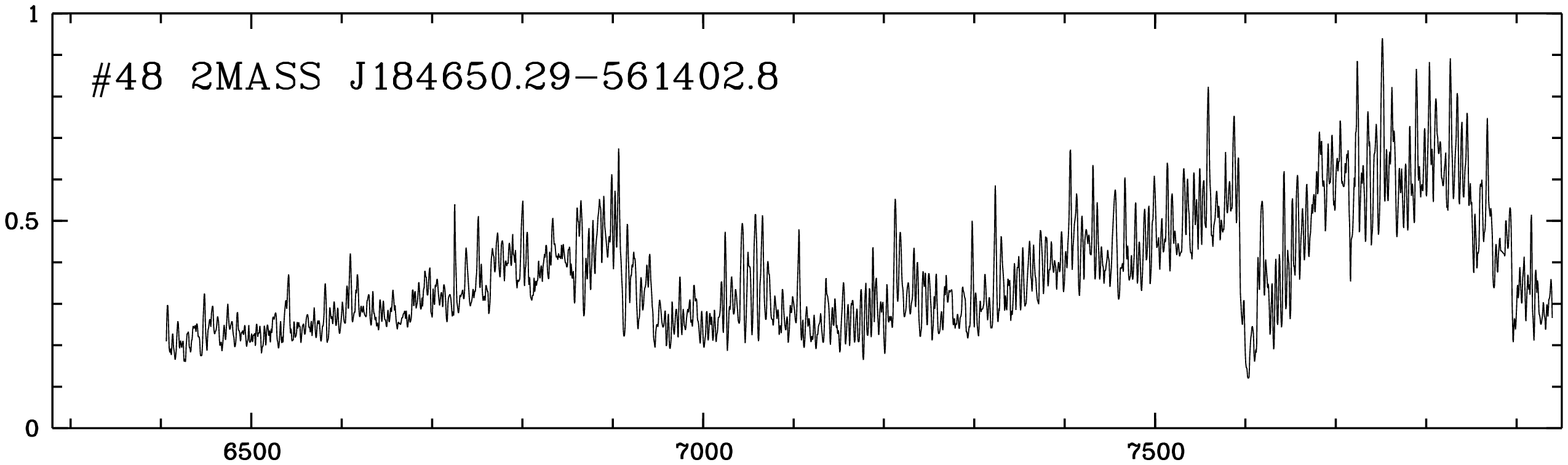}}}
   \resizebox{\hsize}{!}{\rotatebox{-90}{\includegraphics{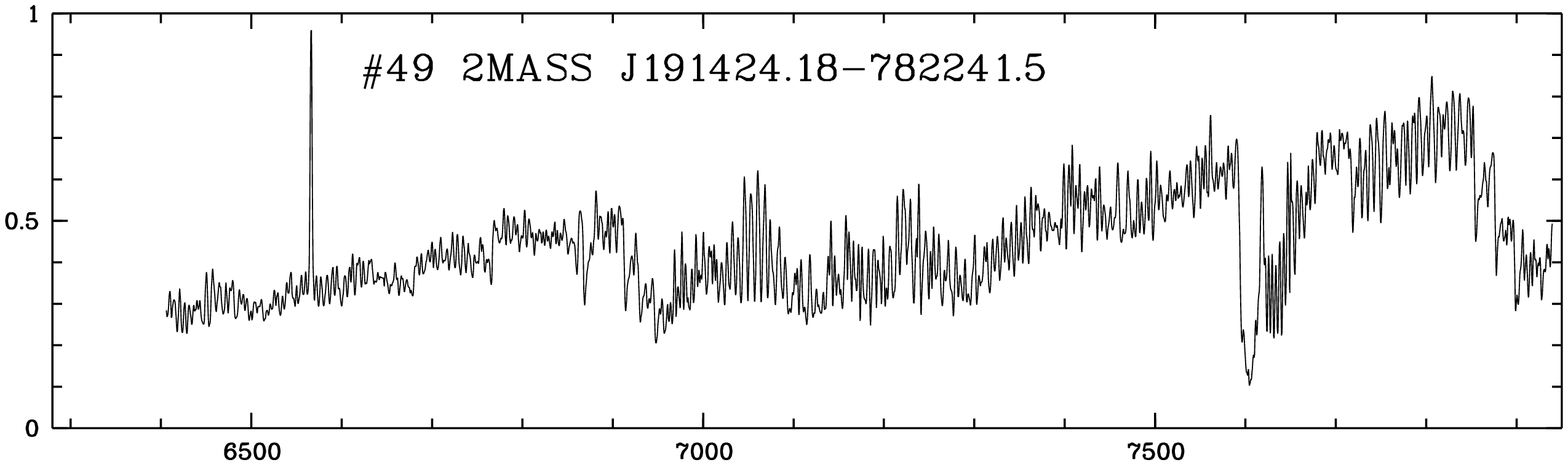}}}
   \resizebox{\hsize}{!}{\rotatebox{-90}{\includegraphics{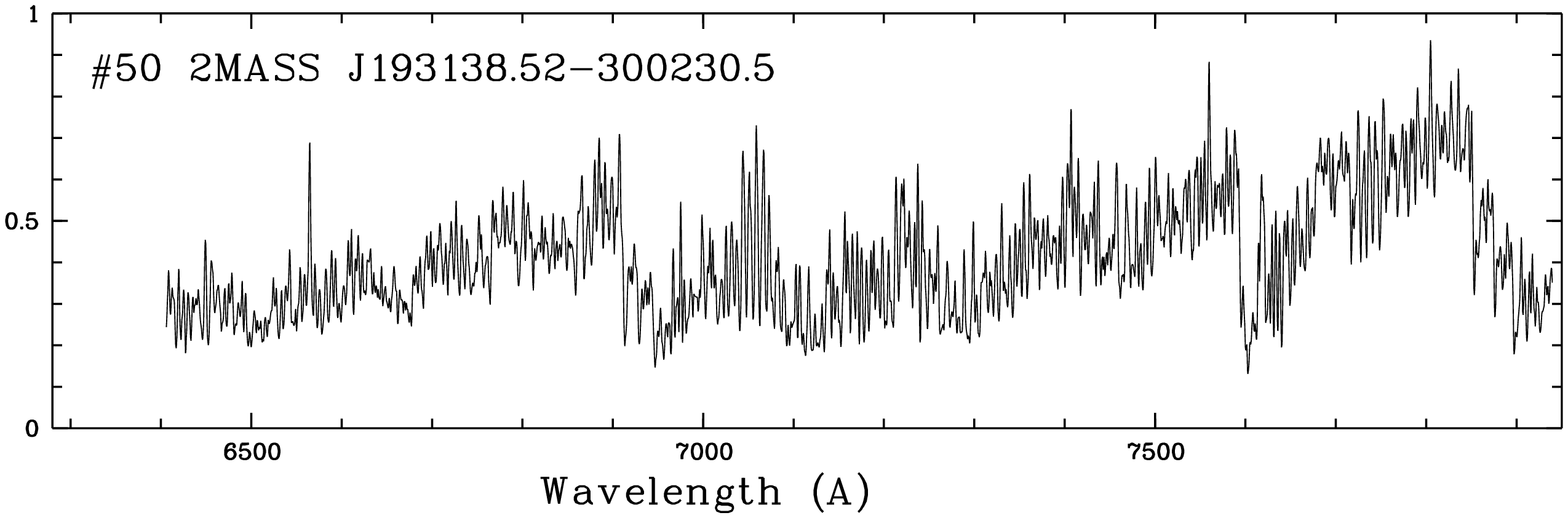}}}

      \caption[]{Spectra of objects 47, 48, 49, \& 50. In all these graphs,
       fluxes in ordinates are in erg\,s$^{-1}$cm$^{-2}$\AA$^{-1}$. Fluxes were divided by factors
       0.15 10$^{-14}$, 2.6 10$^{-15}$,  3.3 10$^{-14}$, \& 0.75 10$^{-14}$
       for objects 47, 48, 49, \& 50, respectively. }
  \label{spe5}
  \end{figure*}
 %===============================================================

   \begin{figure*}

   \resizebox{\hsize}{!}{\rotatebox{-90}{\includegraphics{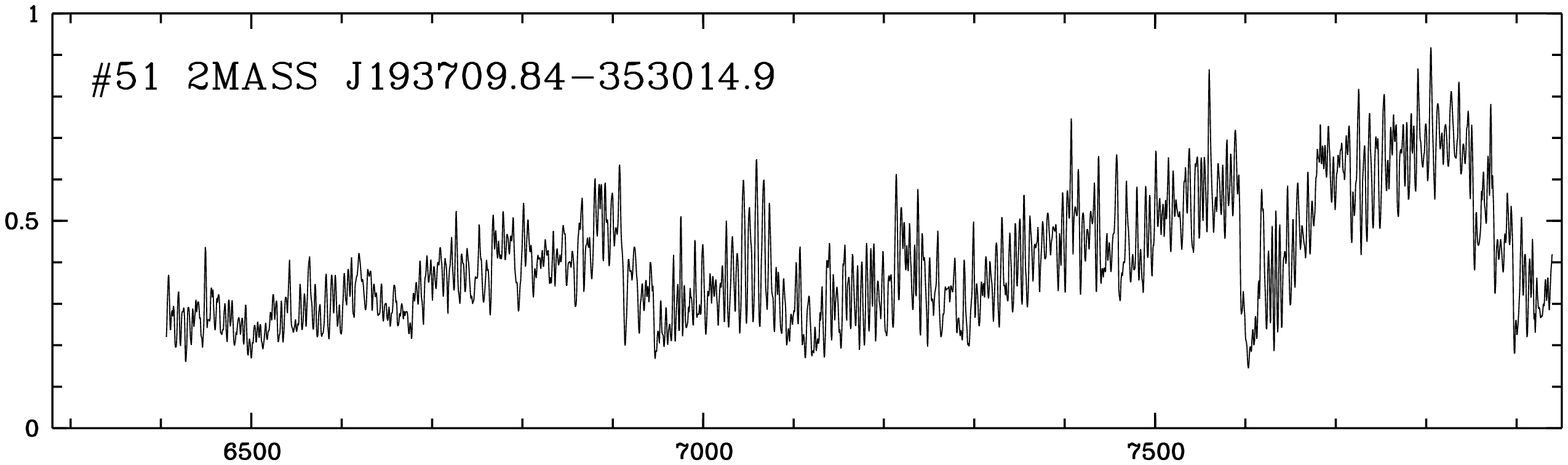}}}
   \resizebox{\hsize}{!}{\rotatebox{-90}{\includegraphics{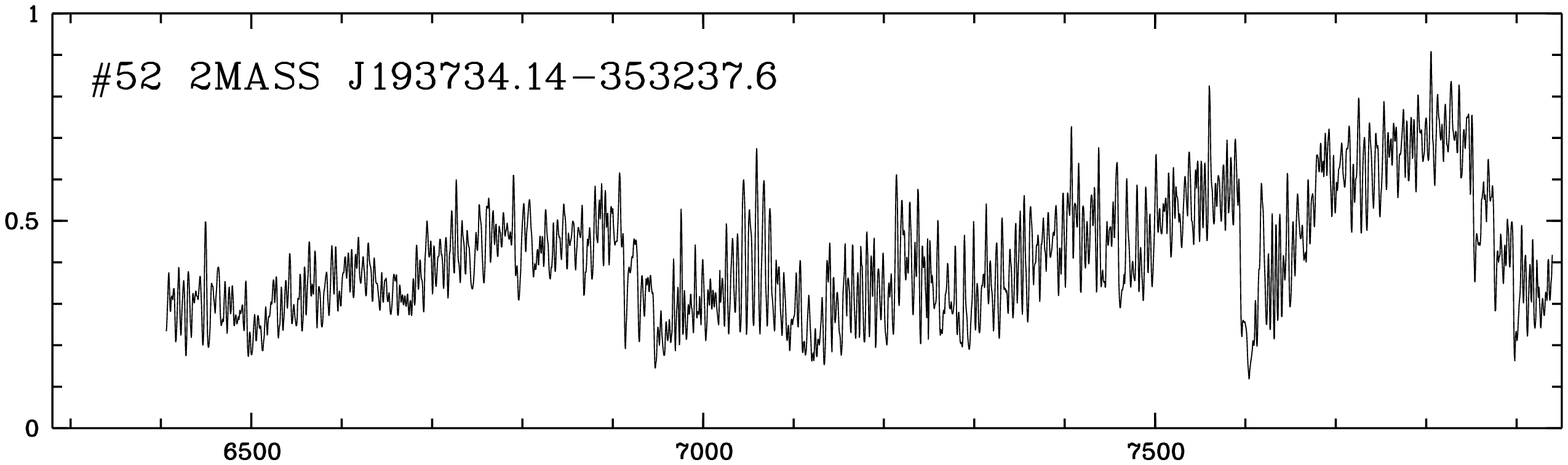}}}
   \resizebox{\hsize}{!}{\rotatebox{-90}{\includegraphics{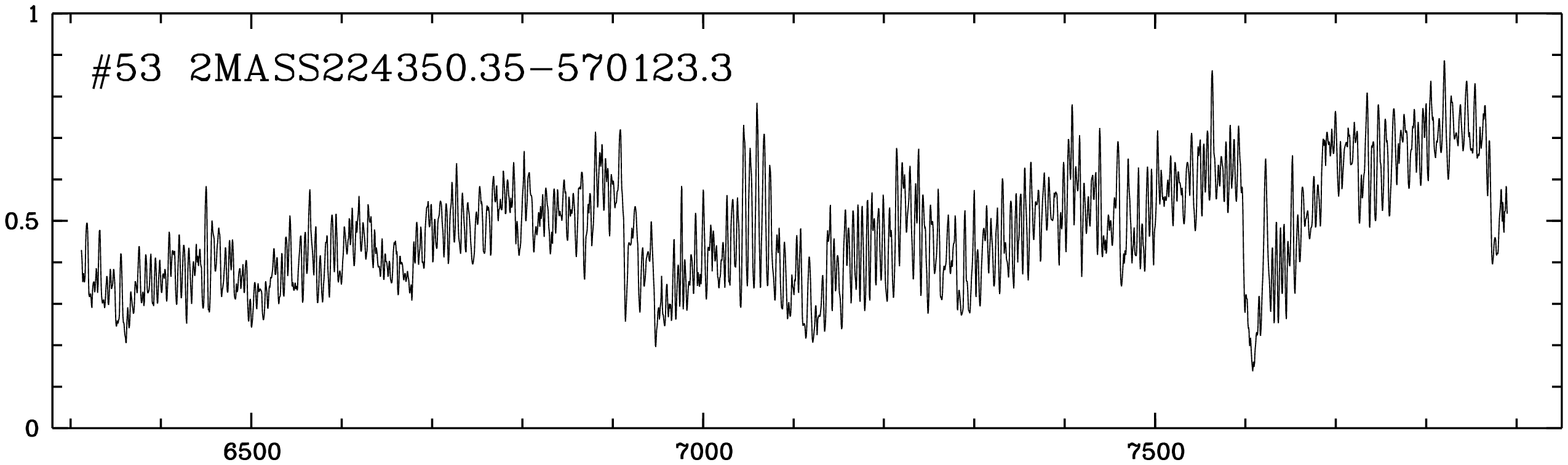}}}
   \resizebox{\hsize}{!}{\rotatebox{-90}{\includegraphics{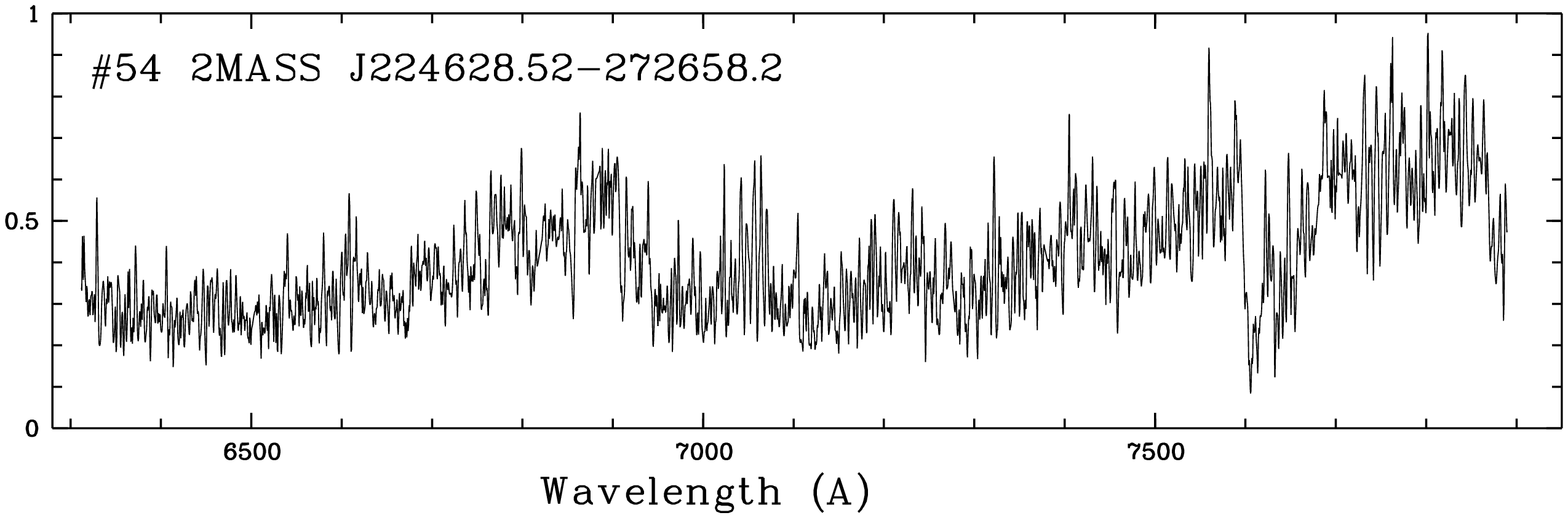}}}

     \caption[]{Spectra of objects 51, 52, 53, \& 54. In all these graphs,
       fluxes in ordinates are in erg\,s$^{-1}$cm$^{-2}$\AA$^{-1}$. Fluxes
       were divided by factors
       0.6 10$^{-14}$, 2.7 10$^{-14}$,  5.5 10$^{-14}$, \& 0.72 10$^{-15}$
       for objects 51, 52, 53, \& 54, respectively. }
   \label{spe6}
   \end{figure*}
 % ========================================================================

    \begin{figure*}
    \resizebox{\hsize}{!}{\rotatebox{-90}{\includegraphics{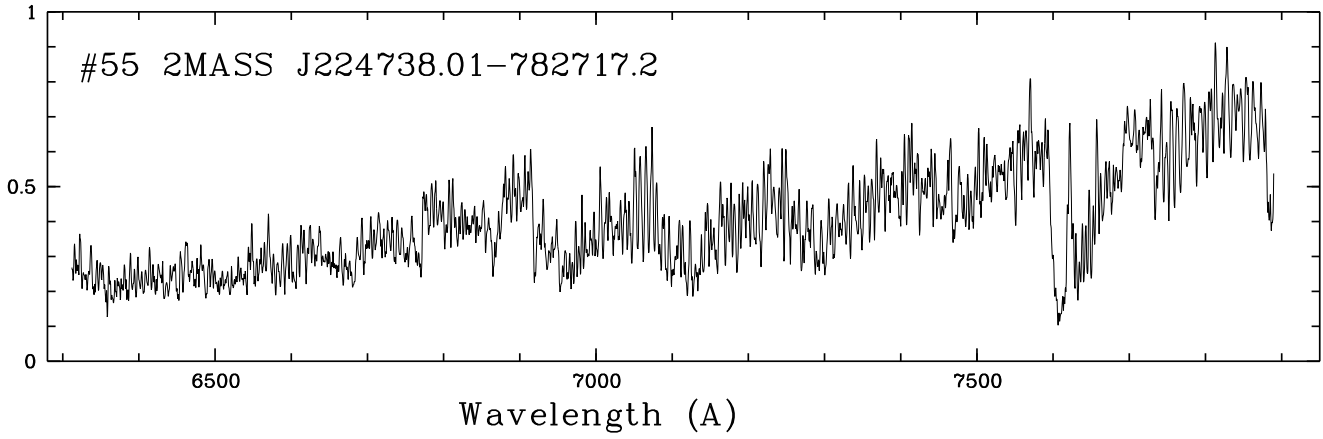}}}

     \caption[]{Spectrum of object 55. The
        flux in ordinates is in erg\,s$^{-1}$cm$^{-2}$\AA$^{-1}$, after dividing by
        a factor of 2.2 10$^{-14}$.}
  \label{spe7}
   \end{figure*}

 % ========================================================================

    \begin{figure*}
    \resizebox{\hsize}{!}{\rotatebox{-90}{\includegraphics{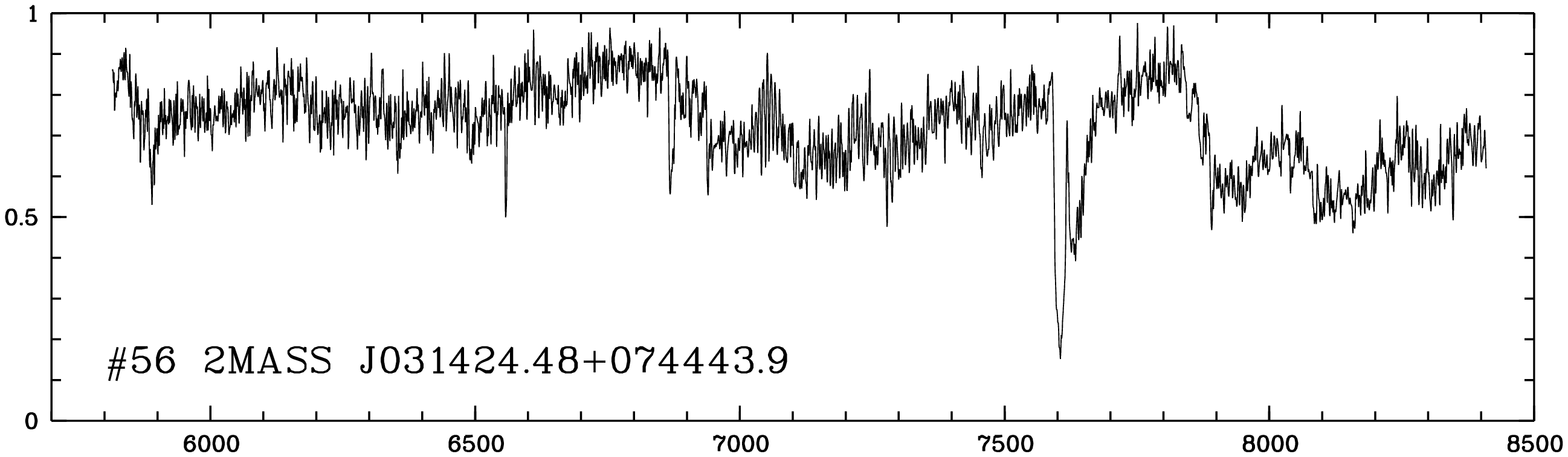}}}
    \resizebox{\hsize}{!}{\rotatebox{-90}{\includegraphics{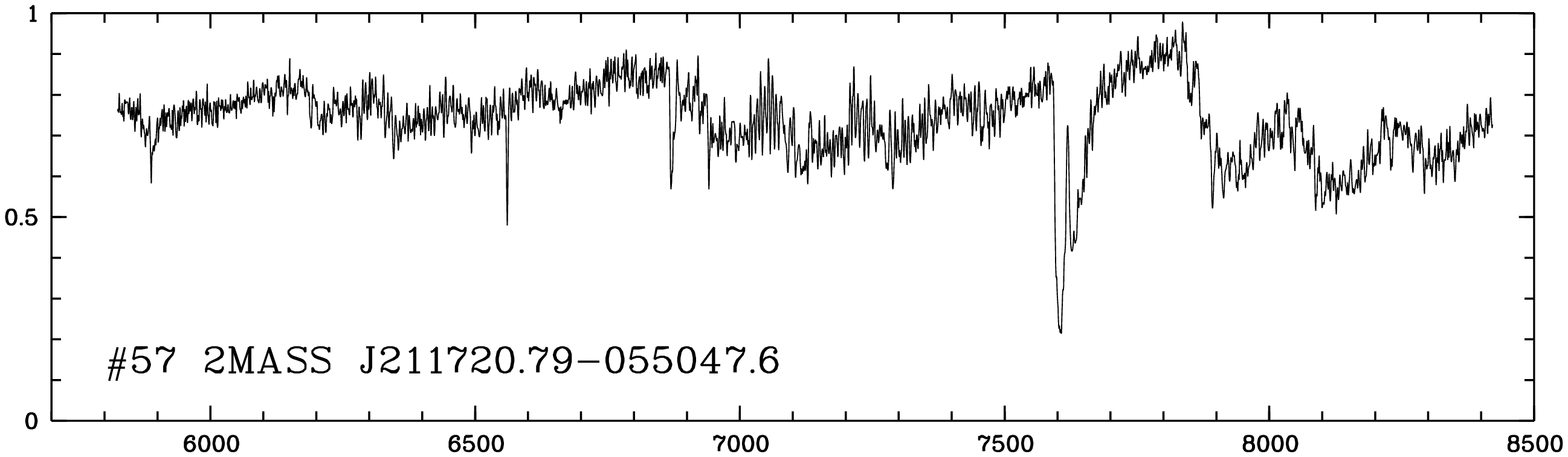}}}
    \resizebox{\hsize}{!}{\rotatebox{-90}{\includegraphics{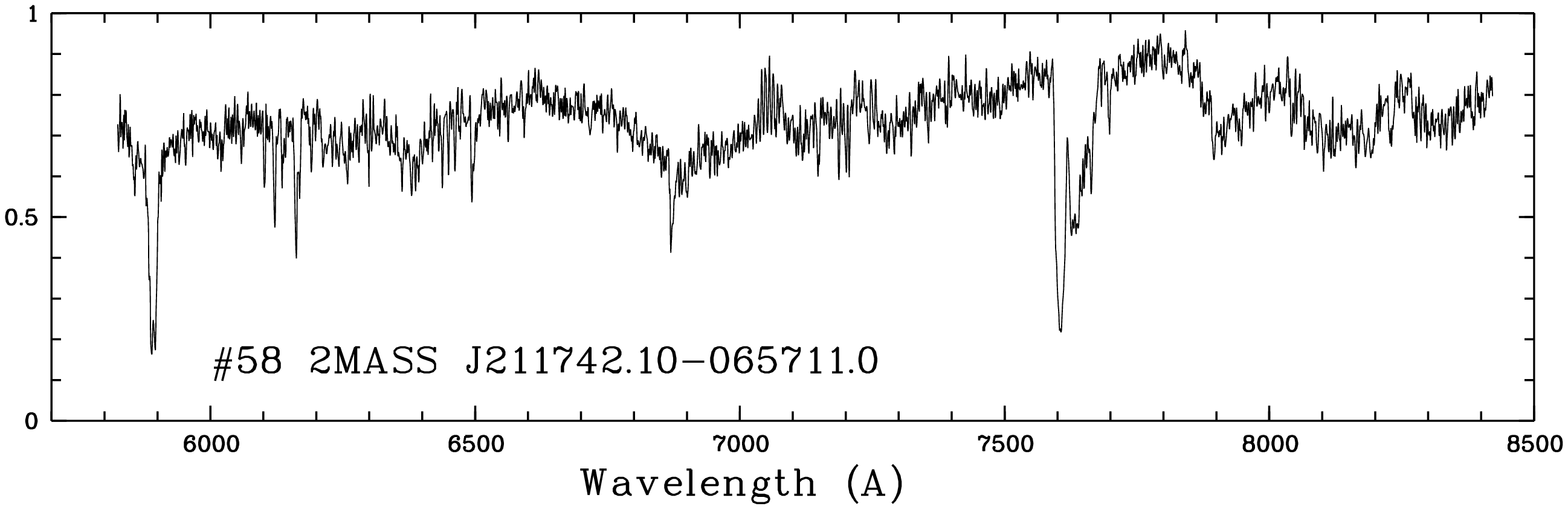}}}
    \caption[]{Spectra of objects 56, 57, \& 58. In all these graphs,
       fluxes in ordinates are in erg\,s$^{-1}$cm$^{-2}$\AA$^{-1}$. 
       Fluxes were divided by factors
      0.21 10$^{-14}$, 0.52 10$^{-14}$, \& 3.7 10$^{-15}$
       for objects 56, 57, \& 58, respectively. }
    \label{spe8}
    \end{figure*}

% ========================================================================

\begin{acknowledgements}

We would like to thank our referee, G. Wallerstein, for many comments that
improved the manuscript.
We acknowledge the use of  the Two Micron All Sky Survey
(2MASS), which is a joint project of the Univ. of Massachusetts and  the Infrared
Processing and Analysis Centre / California Institute of Technology, funded
by NASA and NSF.
This work also benefited from using the CDS database of Strasbourg.
 We also used the POSS-UKST  Digitized Sky Survey made available by
 the Canadian Astronomical Data Centre and by the ESO/ST-ECF Centre in
 Garching. We also acknowledge the use of the MIDAS software of ESO with
 which all data processing and graphs were achieved.

\end{acknowledgements}

%-------------------BIBLIOGRAPHY------------

%\listofobjects
\end{document}